%% file: master.tex
\documentclass{cernrep} 
\usepackage{texnames}
\usepackage[T1]{fontenc}
 \def\bc{\begin{center}}          \def\ec{\end{center}}
\usepackage{subfig}
\usepackage{graphicx}
\usepackage{rotating}

\newcommand{\gsim}{\raisebox{-0.07cm   }
{$\, \stackrel{>}{{\scriptstyle\sim}}\, $}}
\newcommand{\pipi}{\pi^+ \pi^-}
\newcommand{\mpipi}{m_{\pi^+\pi^-}}
\newcommand{\Wgp}{W_{\gamma p}}
\newcommand{\fTwo}{f_{2}}
\newcommand{\gev}{\, \mathrm{GeV}}
\newcommand{\gevsq}{\, \mathrm{GeV}^2}
\newcommand{\xbj}{x}

\begin{document}

\vspace*{-2cm}
\begin{flushright}
DESY--15--253\\
IPPP/15/76\\
DCPT/15/152\\
MAN/HEP/2015/21\\
December 2015
\end{flushright}

\begin{center}
{\Huge \bf Summary of workshop on \\Future Physics with HERA Data}

\vspace{1cm}
{\normalsize \bf A. Bacchetta$^{1}$, J. Bl\"{u}mlein$^{2}$, O. Behnke$^{3}$, J. Dainton$^{4}$, M. Diehl$^{3}$, F. Hautmann$^{5,6}$, A. Geiser$^{3}$, \\
H. Jung$^{3,7}$, U. Karshon$^{8}$, D. Kang$^{9}$, P. Kroll$^{10}$, C. Lee$^{9}$, S. Levonian$^{3}$, A. Levy$^{11}$, \\ 
E. Lohrmann$^{3,12}$, S. Moch$^{12}$, L. Motyka$^{13}$, R. McNulty$^{14}$, V. Myronenko$^{3}$, E.R. Nocera$^{6,15}$, \\ 
S. Pl\"{a}tzer$^{16,17}$, A. Rostomyan$^{3}$, M. Ruspa$^{18}$, M. Sauter$^{19}$, G. Schnell$^{20,21}$, S. Schmitt$^{3}$, \\ 
H. Spiesberger$^{22,23}$, 
I. Stewart$^{24}$, O. Turkot$^{3}$, A. Valk\'arov\'a$^{25}$, K. Wichmann$^{3}$, \\M. Wing$^{26,3,12}$, A.F. \.Zarnecki$^{27}$}

\vspace{1cm}
$^{1}$ University of Pavia and INFN, Pavia, Italy, $^{2}$ DESY, Zeuthen, Germany, $^{3}$ DESY, Hamburg, Germany, \\
$^{4}$ University of Liverpool, UK, $^{5}$ Rutherford Appleton Laboratory, Didcot, UK, \\$^{6}$ University of Oxford, Oxford, UK, 
$^{7}$ University of Antwerp, Antwerp, Belgium, \\$^{8}$ Weizmann Institute, Rehovot, Israel, 
$^{9}$ Los Alamos National Laboratory, Los Alamos, NM, USA, \\
$^{10}$ Bergische Universit\"{a}t Wuppertal, Wuppertal, Germany, 
$^{11}$ Telv Aviv University, Tel Aviv, Israel, \\
$^{12}$ Universit\"{a}t Hamburg, Hamburg, Germany, $^{13}$ Jagiellonian University, Krak\'{o}w, Poland, \\
$^{14}$ University College Dublin, Dublin, Ireland, $^{15}$ University of Genova and INFN Genova, Italy, \\
$^{16}$ Institute for Particle Physics Phenomenology, Durham, UK, \\
$^{17}$ University of Manchester, Manchester, UK, \\ $^{18}$ University of Piemonte Orientale and INFN Torino, Italy, \\
$^{19}$ Universit\"{a}t Heidelberg, Heidelberg, Germany, \\
$^{20}$ University of the Basque Country and Basque Foundation of Science, Bilbao, Spain, \\
$^{21}$ Ghent University, Gent, Belgium, $^{22}$ Johannes Gutenberg-Universit\"{a}t, Mainz, Germany, \\
$^{23}$ University of Cape Town, Rondebosch, South Africa, \\
$^{24}$ Massachusetts Institute of Technology, Cambridge, MA, USA, \\
$^{25}$ Charles University, Praha, Czech Republic, $^{26}$ University College London, London, UK \\
$^{27}$ University of Warsaw, Warsaw, Poland
\end{center}

\vspace{0.cm}
\section*{Abstract}
Recent highlights from the HERA experiments, Hermes, H1 and ZEUS, are reviewed and ideas for future analyses to fully 
exploit this unique data set 
are proposed.  This document is a summary of a workshop on future physics with HERA data held at DESY, Hamburg at the 
end of 2014.  All areas of HERA physics are covered and contributions from both experimentalists and theorists are included.  
The document outlines areas where HERA physics can still make a significant contribution, principally in a deeper understanding of 
QCD, and its relevance to other facilities.  Within the framework of the Data Preservation in High Energy Physics, the HERA data have 
been preserved for analyses to take place over a timescale of 10 years and more.  Therefore, although an extensive list of 
possibilities is presented here, safe storage of the data ensures that it can also be used in the far future should new ideas and analyses 
be proposed.

\pagebreak[4]
\
\tableofcontents
\pagebreak[4]
\


 \input{executive}

 
 \input{highlights}


\input{general}


\input{pdfs}


\input{jets}


\input{common}


\input{diffraction}


\input{spin}


\input{montecarlo}


\input{summary}

\input{acknowledgements}

\pagebreak[4]
\


\input{bib}
\end{document}

%% file: executive.tex
 \section{Introduction}

These following proceedings are summaries from individual speakers and others who contributed to the symposium and workshop 
on future physics with HERA data.   In the symposium, some of the latest experimental results from the HERA collaborations are reviewed 
and a theoretical perspective is given (see Section~\ref{sec:symp}).  The workshop on future physics ideas was organised thematically 
into sessions which are then organised into the following sections in this write-up:

\begin{itemize}

\item Overview of perspectives on physics with HERA data (see Section~\ref{sec:general});

\item Parton density functions and electroweak effects (see Section~\ref{sec:pdfs});

\item Jets and hadronic final states (see Section~\ref{sec:jets});

\item Physics topics common with other experiments (see Section~\ref{sec:common});

\item Diffraction and low-$x$ physics (see Section~\ref{sec:diffraction});

\item Spin physics (see Section~\ref{sec:spin});

\item Monte Carlo programmes for HERA physics (see Section~\ref{sec:montecarlo});

\end{itemize}

Additionally, a technical presentation was given~\cite{dirk-dphep} on the safe storage and maintenance of the HERA data for future 
analyses for many years to come, as part of the global DPHEP effort~\cite{1512.02019}.  Finally a summary of the workshop was 
given (see Section~\ref{sec:summary}).

%% file: highlights.tex
\section{Recent highlights from HERA}
\label{sec:symp}

\input{Contributions/AharonLevy/levy.tex}

\input{Contributions/AliceValkarova/valkarova.tex}

%% file: Contributions/AharonLevy/levy.tex
\subsection{Recent HERA results on proton structure\\
{\it A. Levy}
}

The latest results on the proton structure by the two HERA collaborations, H1 and ZEUS, are presented. They include new results since the last HERA Symposium and cover: the cross section data combinations; the new HERAPDF2.0 parton distribution functions; measurements of the charm and bottom structure functions, with the determination of the masses of these heavy quarks; energy dependence of $D^*$-meson production; high-$Q^2$ measurements from the ZEUS collaboration in the high-Bjorken-$x$ region up to values of $x\cong 1$; and longitudinal structure function $F_L$ from both collaborations. 

Though the HERA collider stopped running in 2007, the data analysis has been going on with big efforts on finalising the inclusive cross section measurements. As both collaborations have done so~\cite{zfinal,H1final}, the way was paved for the `climax' of an experiment; the data combination. This procedure, which had produced precision measurements already for the HERA\ I period of running~\cite{HERA1-comb}, has produced text-book results of the HERA cross section measurements. In the new combination, there were 41 data sets including 2927 data points which were combined to 1307 averaged measurements with 165 sources of correlated systematic uncertainties. The data sets were consistent, producing a total $\chi^2$/ndf = 1.04.

\begin{figure}[h!]
\begin{minipage}{0.52\linewidth}
\includegraphics[width=0.9\linewidth]{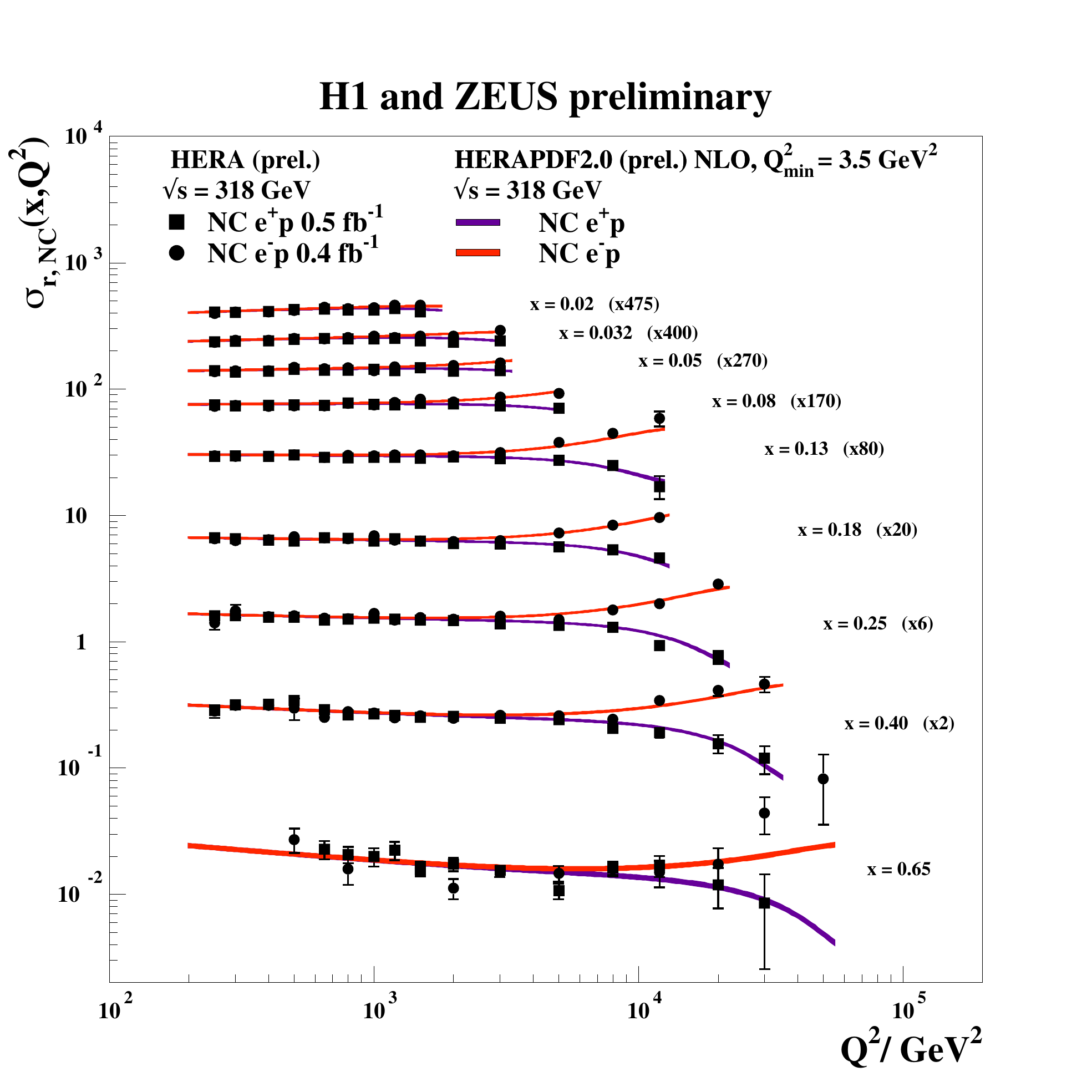} 
\caption{The combined HERA data for the inclusive NC $e^+ p$ and $e^- p$ reduced cross sections as a function of $Q^2$ for selected values of $x$ at $\sqrt{s}$ = 318\,GeV with overlaid predictions of the preliminary version of HERAPDF2.0 NLO. The bands represent the total uncertainties of the predictions.}
\label{fig:nc-comb} 
\end{minipage}
\hfill
\begin{minipage}{0.46\linewidth}
\includegraphics[width=0.9\linewidth]{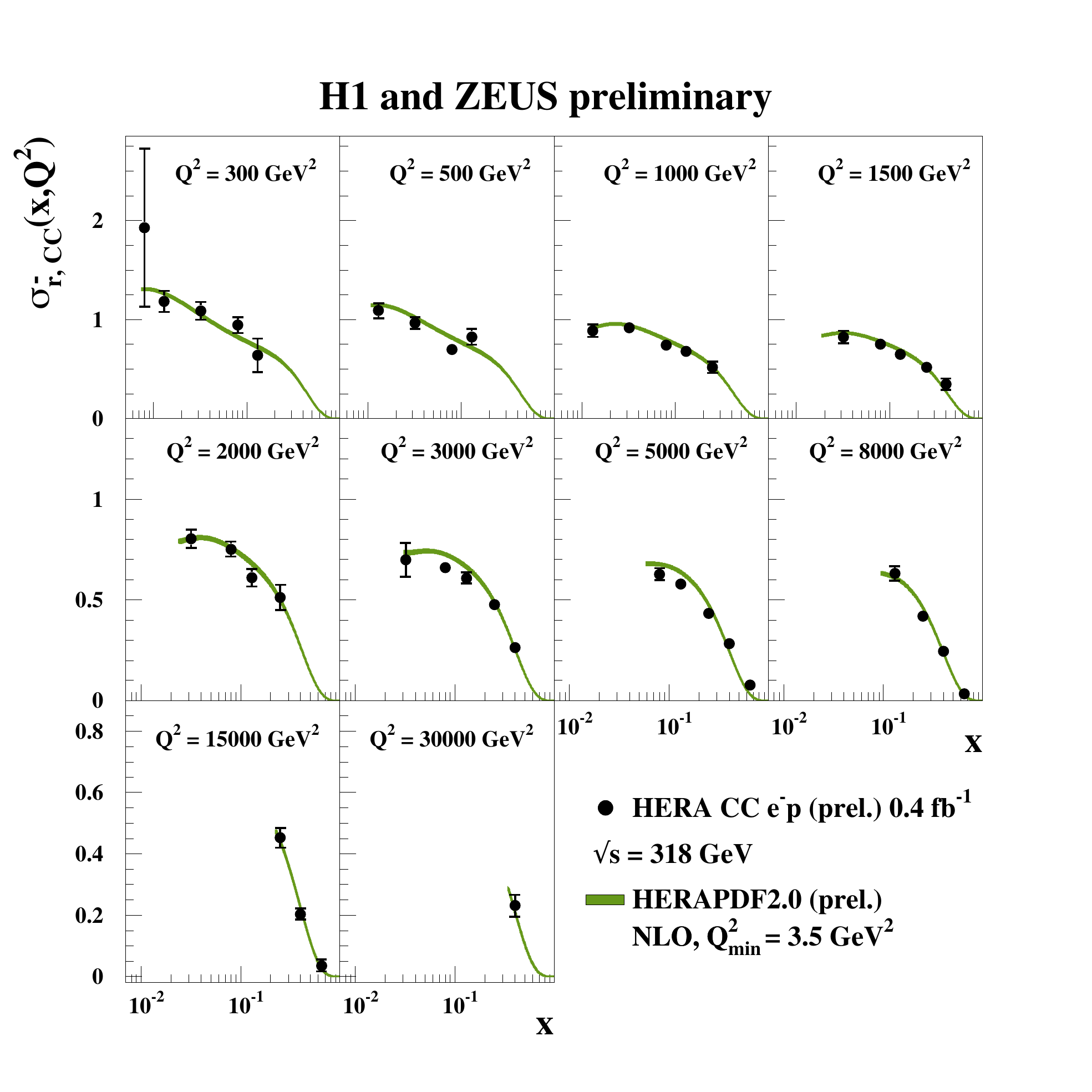} 
\vspace{5mm}
\caption{The combined HERA inclusive CC $e^-ˆ' p$ reduced cross sections as a function of $x$ at selected values of $Q^2$ $\sqrt{s}$ = 318\,GeV with overlaid predictions of the preliminary version of HERAPDF2.0 NLO. The bands represent the total uncertainties on the predictions.} 
\label{fig:cc-comb} 
\end{minipage}
\end{figure}

The scaling violation plot, namely the combined neutral current (NC) $e^+ p$ and $e^-' p$ reduced cross sections as a function of $Q^2$ for fixed values of Bjorken $x$, for all HERA data resulting from an integrated luminosity of 
1\,fb$^{-1}$ is shown in Fig.~\ref{fig:nc-comb}. They are the most precise data measurements in this kinematic region, reaching $Q^2$ values of 50 000\,GeV$^2$ and $x$ values down to 0.00005 (in the present figure only values down to 0.02 are shown). The scaling violations are well described by the HERAPDF2.0 NLO predictions.

Figure~\ref{fig:cc-comb} shows the combined HERA charged current (CC) $e^- p$  reduced cross sections as a function of $x$ for  some fixed $Q^2$ values as indicated in the figure. These data are also well described by the HERAPDF2.0 NLO predictions.   Figures~\ref{fig:nc-comb} and~\ref{fig:cc-comb} are part of the HERA legacy. They are the results of many years of dedicated work to produce the best quality data by the two HERA collaborations H1 and ZEUS.

\begin{figure}[h!]
\begin{minipage}{0.4\linewidth}
\centerline{\includegraphics[width=0.99\linewidth]{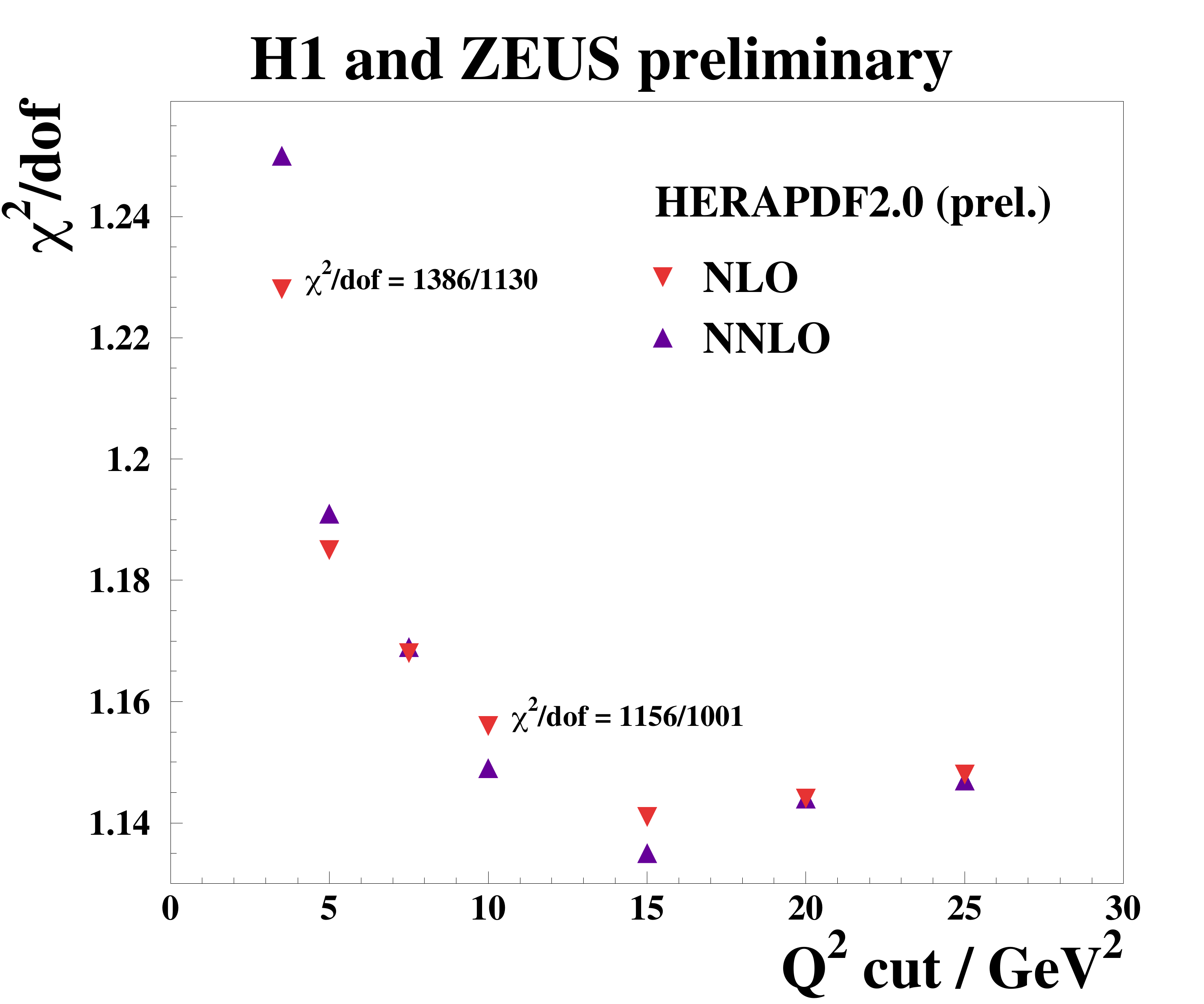}}
\caption{The dependence of $\chi^2$/dof on $Q^2_{\rm cut}$ of the NLO and NNLO fits to the HERA combined inclusive data.}
\label{fig:chisq}
\end{minipage}
\hfill
\begin{minipage}{0.58\linewidth}
The combined data are used to obtain the parton distribution function (PDFs) in the proton, using the DGLAP~\cite{dglap} equations. Since perturbative QCD (pQCD) is expected to be valid only from a scale where partons can be resolved, the pQCD analysis uses data above that scale. A lower value for  $Q^2_{\rm cut}$ was chosen as 3.5\,GeV$^2$ and a study of the resulting  $\chi^2$/dof as a function of this cut was carried out and shown in Fig.~\ref{fig:chisq}. A clear dependence of the $\chi^2$/dof on $Q^2$ is observed; its value decreases as $Q^2$ increases to $\sim$ 10\,GeV$^2$ and starts to increase again after 15\,GeV$^2$. Following this behaviour, two sets of PDFs have been determined, one with $Q^2_{\rm min}$ = 3.5\,GeV$^2$ and another with $Q^2_{\rm min}$ = 10\,GeV$^2$.
\end{minipage}
\end{figure}

\begin{figure}[h!]
\begin{minipage}{0.49\linewidth}
\includegraphics[width=0.98\linewidth]{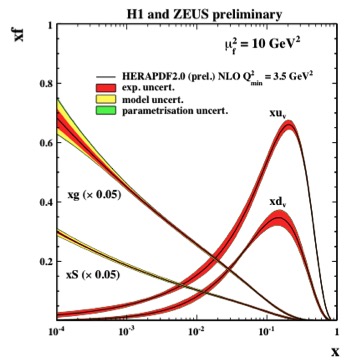} 
\caption{The parton distribution functions $xu_v$, $xd_v$, $xS = 2x(\bar{U} + \bar{D})$ and $xg$ of HERAPDF2.0 NLO at $\mu^2_f$ =
3.5\,GeV$^2$. The gluon and sea distributions are scaled down by a factor of 20. The experimental, model and parameterisation uncertainties are shown. }
\label{fig:qmin3.5} 
\end{minipage}
\hfill
\begin{minipage}{0.49\linewidth}
\includegraphics[width=0.98\linewidth]{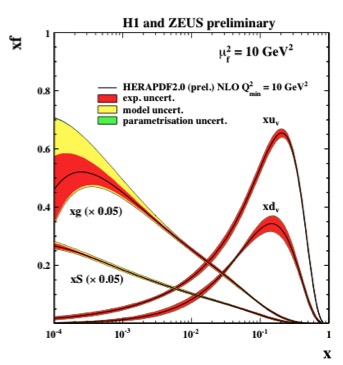} 
\caption{The parton distribution functions $xu_v$, $xd_v$, $xS = 2x(\bar{U} + \bar{D})$ and $xg$ of HERAPDF2.0 NLO at $\mu^2_f$ = 
10\,GeV$^2$. The gluon and sea distributions are scaled down by a factor of 20. The experimental, model and parameterisation uncertainties are shown.} 
\label{fig:qmin10} 
\end{minipage}
\end{figure}

The two sets of PDFs are shown in Figs.~\ref{fig:qmin3.5} and \ref{fig:qmin10}. Although there is no visible difference between the two in the higher-$x$ region, one sees that the gluon uncertainty is much larger in the low-$x$ region for the $Q^2_{\rm min} = $ 10\,GeV$^2$ case. This happens because leaving out data in the range $Q^2$ = 3.5 -- 10\,GeV$^2$ removes events which constrain the gluon density in the low-$x$ region. Further details and final results on the combined inclusive HERA data can be found in the recent publication~\cite{combEPJC2015}.

H1 and ZEUS combined their charm structure function data and extracted the running charm mass~\cite{mc}, resulting in $m_c(m_c) = 1.26 \pm 0.05 ({\rm exp}) \pm 0.03 ({\rm mod}) \pm 0.02 ({\rm param}) \pm 0.02 ({\alpha_S})$\,GeV. The ZEUS collaboration measured the bottom structure function~\cite{mb} and extracted the $\overline{\rm MS}$ value of the bottom quark, $m_b(m_b) = 4.07 \pm 0.14 ({\rm exp}) ^{+0.01}_{-0.07}  ({\rm mod}) ^{+0.05}_{-0.00} ({\rm param})  ^{+0.08}_{-0.05} ({\rm theo})$\,GeV.

The ZEUS collaboration measured the energy dependence of the $D^*$ cross section~\cite{dstar} in the range of $\sqrt{s}$ = 225 - 320 GeV and found the dependence to be in agreement with NLO QCD predictions. It also used an improved reconstruction method of Bjorken $x$ in the high-$Q^2$ and high-$x$ region and extracted~\cite{zeushighx} the inclusive NC cross section up to $x \to$ 1.  The fine binning in $x$ with the extension of kinematic coverage up to $x \cong$ 1 make these data important input to fits constraining the PDFs in the valence-quark domain.

\begin{figure}[h!]
\begin{minipage}{0.43\linewidth}
The H1 collaboration measured~\cite{h1fl} the $F_L$ structure function in the region $1.5 < Q^2 < 800$ GeV$^2$ while the  ZEUS collaboration measured it~\cite{zeusfl} in the kinematic range $9 < Q^2 < 110$ GeV$^2$. The results are shown in Fig.~\ref{fig:fl-both}. The uncertainties of the ZEUS results are larger than those of H1. The ZEUS results, though consistently lower than those of H1, are consistent with them because of the correlated uncertainties. The predictions shown by the shaded area are in reasonable agreement with both data sets. One would hope to have a combined measurement which will produce the HERA $F_L$.
\end{minipage}
\hfill
\begin{minipage}{0.55\linewidth}
\includegraphics[width=0.98\linewidth]{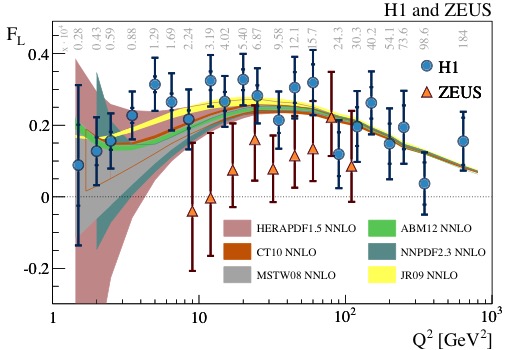} 
\caption{$F_L$ as a function of $Q^2$ as measured by the H1 and ZEUS collaborations. The shaded area are predictions based on different PDF parameterisations, as indicated in the figure.} 
\label{fig:fl-both} 
\end{minipage}
\end{figure}

%% file: Contributions/AliceValkarova/valkarova.tex
\subsection{Results on the hadronic final state and diffraction in $ep$ scattering\\
{\it A. Valk\'arov\'a}
}

\subsubsection{QCD and hadronic final state}\label{aba:sec1}

The ZEUS collaboration presented cross sections for events containing an isolated high-energy photon, with and without a 
jet, produced in photoproduction using the full HERA II data set \cite{photons1}. These measurements were later on 
extended~\cite{photons2}.  Events with isolated photons can provide a direct probe of the underlying partonic process in 
high-energy collisions, since the emission of a high-energy photon is not affected by hadronisation.  Within the large theoretical 
uncertainties, the theoretical predictions agree with the data.

Inclusive jet, dijet and trijet differential cross sections were measured in neutral current deep-inelastic scattering (DIS) collisions 
using the H1 detector~\cite{jets}.  Theoretical QCD calculations at NLO, corrected for hadronisation and electroweak effects, 
provide a good description of the measured single- and double-differential jet cross sections as a function of all studied variables. 
This measurement was confirmed also by a ZEUS analysis measuring trijet differential cross sections~\cite{trijets}. The H1 collaboration 
derived the most precise value of the strong coupling constant from jet data at NLO as measured in one experiment,  
$\alpha_s(M_Z) = 0.1165~(8)_{\rm exp}~(38)_{\rm pdf,theo}$.
  
A sophisticated analysis of H1 data aimed to find instantons in DIS $ep$ interactions\cite{instantons}. The observed upper limit on 
the QCD instanton cross section of about 1.6\,pb  does not support the theoretical prediction. This analysis is still continuing.

\subsubsection{Diffraction}

\begin{figure}
\begin{center}
\resizebox{0.5\textwidth}{!}{%
  \includegraphics{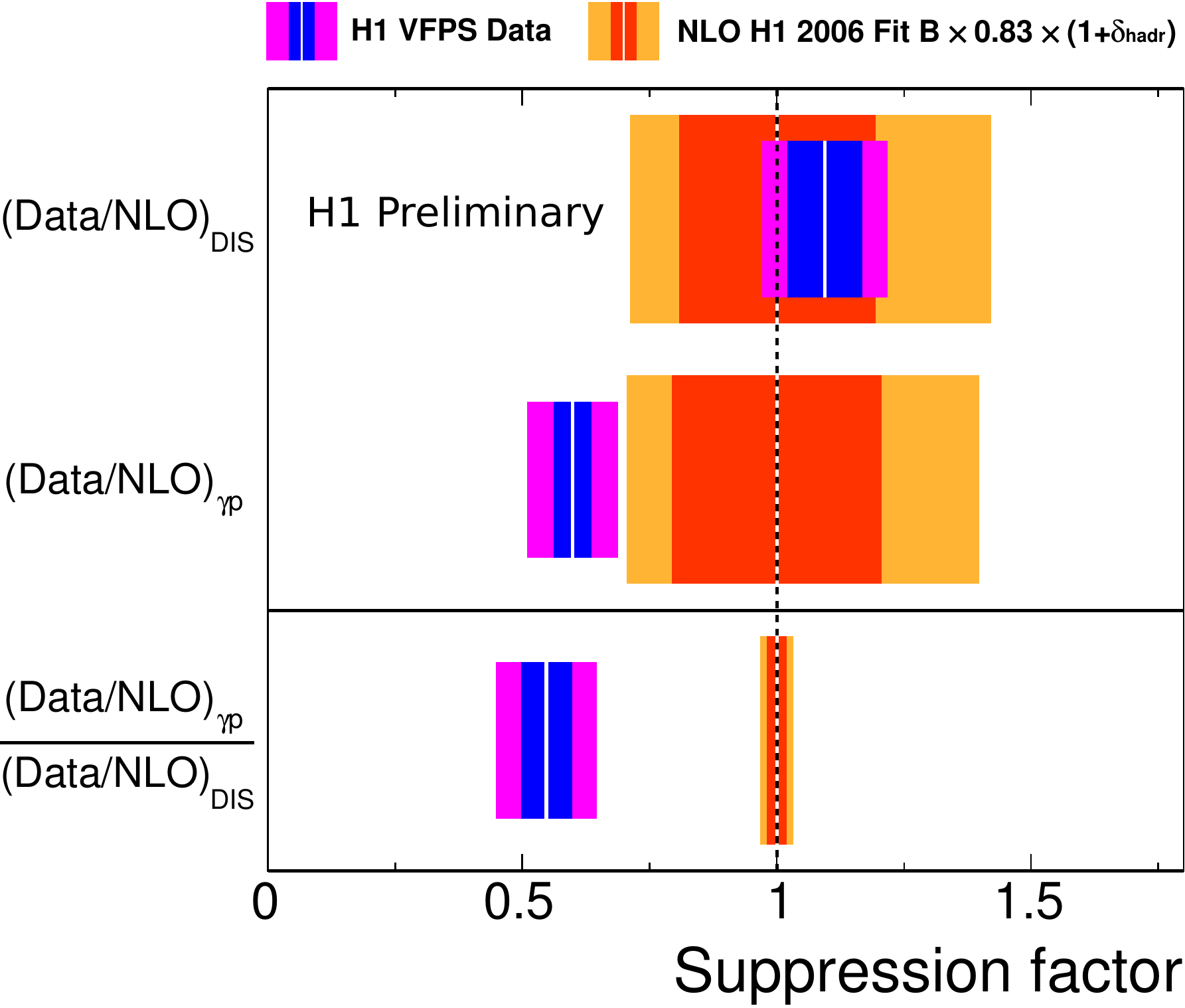}
}
\caption{DIS and photoproduction integrated cross sections normalised to the NLO QCD theoretical
calculations are shown as a white line. The double ratio of data over NLO QCD calculations for photoproduction
to data over NLO QCD calculations for DIS is presented as a white line in the last row.}
\label{fig:doub}
\end{center}
\end{figure}

It has been established by the ZEUS collaboration that the measurement of the shape of the azimuthal angular distributions 
of exclusive dijets in diffractive DIS prefers a 2-gluon exchange model of $q\overline{q}$ production over resolved Pomeron 
model \cite{angle}.
     
The H1 collaboration presented the single- and double-differential dijet cross sections in diffractive DIS $ep$ scattering using 
the large rapidity gap method to select diffractive events.  Both shapes and normalisation of the single-differential cross sections 
are reproduced satisfactorily by the NLO QCD predictions within the experimental and theory uncertainties~\cite{Boris}.

A new measurement of the diffractive dijet cross section with a leading final state proton detected in the Very Forward Proton Spectrometer 
was performed by the H1 collaboration in photoproduction and DIS~\cite{Radek}. With the exception of the momentum transfer $Q^2$, 
identical kinematic regions were used for DIS and photoproduction dijets.  In both photoproduction and DIS the shapes of differential 
distributions are well described by a NLO QCD calculations. However,  for photoproduction the theoretical predictions lie systematically 
above the data, corresponding to a data suppression factor of about 0.6. For DIS, the NLO QCD calculations agree within theoretical 
uncertainties with data.
   
Integrated over the measured kinematic range, the double ratio of data over NLO QCD calculations for photoproduction to data over 
NLO calculations for DIS is:
\begin{equation}
\frac{(\mathrm{DATA}/\mathrm{NLO})_{\gamma p}}{(\mathrm{DATA}/\mathrm{NLO})_{\mathrm{DIS}}} =0.55\pm0.10\,(\mathrm{data}) \pm0.02\,(\mathrm{theo.})
\end{equation}
and is shown in Fig.~\ref{fig:doub}. This observation is in agreement with previous H1 measurements, where complementary experimental 
methods have been used. Within the theoretical framework applied, this could hint at the breaking of QCD factorisation in diffractive dijet 
photoproduction.

%% file: general.tex
\section{Overview of perspectives on physics with HERA data}
\label{sec:general}

\input{Contributions/HannesJung/jung.tex}

\input{Contributions/AchimGeiser/geiser.tex}

%% file: Contributions/HannesJung/jung.tex
\subsection{Transverse momentum dependent parton distributions\\
{\it H.~Jung, F.~Hautmann}
}
The calculation of a cross section in $ep$ and $pp$ collisions proceeds via the convolution of a partonic cross section with parton density functions 
(PDFs), which describe the probability density to find a parton in a hadron with a given momentum fraction $x$ at a resolution scale $\mu$. For 
inclusive single scale cross sections, like the inclusive DIS cross section, the parton densities are functions of $x$ and the resolution scale 
$\mu$ only (collinear factorisation), while for more exclusive and differential multi-scale cross sections, the transverse momenta of the interacting 
partons become important, and so-called transverse momentum dependent (TMD) parton density functions are needed. 

DIS is the process where TMD factorisation is proven. The precision determination of TMDs (for quarks and gluons) is extremely important especially 
in light of potential factorisation breaking effects in hadron--hadron collisions~\cite{Collins:2007nk,Catani:2014qha}.  
The present status of TMDs is summarised in~\cite{Angeles-Martinez:2015sea}; a library of available TMD fits and parameterisations can be found in~\cite{Hautmann:2014kza};  an application of TMDs to $W$+jet production in $pp$ is given in~\cite{Dooling:2014kia}.

The TMDs, similarly to collinear PDFs, can be determined from fits to inclusive cross sections~\cite{Hautmann:2013tba}, but the transverse momentum 
dependence is, at present, not well constrained. Dedicated measurements are needed to constrain the small and large $k_T$ region of quark and gluon 
TMDs:
\begin{itemize}
\item the inclusive jet cross section from lowest $p_T\sim {\cal O}(0.5\,{\rm GeV})$ to large $p_T$ to constrain the quark and gluon TMDs

\item inclusive particle- and identified particle-production as a function of $x$, $Q^2$, $\eta$ and $p_T$  of the tracks to constrain the flavour dependence 
of the TMDs
\end{itemize}

In jet production in $\gamma p$, but also in DIS, the region transverse in azimuth to the jet direction is usually not well described by standard simulations 
including parton showers (underlying event measurements), and especially for $\gamma p$, multiple parton interactions (MPIs) must be included, processes 
which are also needed in $pp$. However, the separation of multiple interactions and higher order contributions from a single interaction depends in general 
on the factorisation used and on the factorisation scheme. Especially measurements in DIS are crucial, since there MPIs are not expected to contribute because 
of the point-like nature of the virtual photon. Such measurements would be important to constrain the contribution from higher order QCD radiation in a single 
partonic interaction. With this, a comparison with $\gamma p$ or $pp$ would allow a systematic estimate of the MPI contribution, in a factorisation scheme 
independent way.

%% file: Contributions/AchimGeiser/geiser.tex
\subsection{Possible future HERA analyses\\
{\it A. Geiser}}

The purpose of this contribution is to give a (subjective)  overview of a large variety of analyses which might be possible in the future using 
the HERA data. It is based on 
\begin{itemize}
\item a comparison of early goals with what has actually been achieved,
\item an extrapolation of how already existing results can be improved further, and
\item a collection of thoughts about analysis topics which might never have been attempted before.  
\end{itemize}
Because there is obviously large overlap, many of the ideas presented by other participants of the workshop are integrated as 
much as possible and need to be appropriately quoted and referenced.
Doing this adequately is not possible within the limited space available for a single contribution to this document.
Therefore the bulk of the information, with the corresponding references, is presented in a separate document~\cite{possibleHERA},
which is only briefly summarised here.

A variety of possible future analyses of HERA data in the context of the DPHEP data preservation programme~\cite{1512.02019,DPHEP} is collected, motivated, and 
commented. The focus is placed on possible future analyses of the existing $ep$ collider data and their physics scope. Comparisons to the original 
scope of the HERA programme are made, and cross references to topics also covered by other participants of the workshop are given.
This includes topics on 
\begin{itemize}
\item QCD, 
\item proton structure, 
\item diffraction, 
\item jets, 
\item hadronic final states, 
\item heavy flavours, 
\item electroweak physics,
\end{itemize} 
and the application of related theory and phenomenology topics like 
\begin{itemize}
\item NNLO QCD calculations, 
\item low-$x$ related models, 
\item nonperturbative QCD aspects, and 
\item electroweak radiative corrections.  
\end{itemize}
Synergies with other collider programmes like LHC and EIC are also addressed.

In summary, the range of physics topics which can still be uniquely covered using the existing data is very broad and of considerable 
physics interest, often matching the interest of results from colliders currently in operation. Due to well-established 
data and MC sets, calibrations, and analysis procedures the manpower and expertise needed for a particular analysis is often very much
smaller than that needed for an ongoing experiment. Since centrally funded manpower to carry out such analyses is not available any
longer, this contribution, and in particular its extended version~\cite{possibleHERA}, not only addresses experienced self-funded experimentalists, but 
also theorists and master-level students who might wish to carry out such an analysis.

%% file: pdfs.tex
\section{Parton density functions and electroweak physics}
\label{sec:pdfs}

\input{Contributions/JohannesBluemlein/bluemlein}

\input{Contributions/SvenMoch/moch}

\input{Contributions/HubertSpiesberger/spiesberger}

\input{Contributions/KatarzynaWichmann/wichmann}

%% file: Contributions/JohannesBluemlein/bluemlein.tex
\subsection{Three-loop heavy flavour corrections to deep-inelastic scattering\\
{\it J. Bl\"umlein}
}

The present precision of the deep-inelastic world data requires QCD analyses and fits to the heavy quark masses $m_c$ and $m_b$
at 3-loop order \cite{Bethke:2011tr,Moch:2014tta,Alekhin:2012vu}. While the massless corrections to 3-loop order are available 
\cite{Moch:2004pa,Vogt:2004mw,Vermaseren:2005qc}, the heavy flavour corrections are being calculated. Here, emphasis is placed  
on an analytic calculation in the region $Q^2 \gg m_Q^2$, which holds in the case of the structure function $F_2(x,Q^2)$
at the 1\% level, if $Q^2/m^2 \gsim 10$ and is a very good approximation in the case of charm in the kinematic regime at HERA \cite{Buza:1995ie}.

In 2009, the programme could be accomplished completely on the level of a series Mellin moments \cite{Bierenbaum:2009mv} in calculating projected 
tadpoles using {\tt MATAD} \cite{Steinhauser:2000ry}. In the meanwhile also moments for a new class of 3-loop diagrams, containing both 
charm and bottom corrections have been computed for the moments $N = 2,4,6$ \cite{BW1}. Very recently all logarithmic corrections were
computed for the Wilson coefficients contributing to $F_2(x,Q^2)$ and all transition matrix elements in the variable flavor number scheme (VFNS)
to 3-loop order \cite{Behring:2014eya}. Here also the first two  Wilson coefficients contributing with  3-loop order, $L_g^S(x,Q^2)$ and 
$L_q^{PS}(x,Q^2)$, were calculated.
The asymptotic 3-loop corrections in the case of the longitudinal structure function $F_L(x,Q^2)$ are also known
\cite{Behring:2014eya,Blumlein:2006mh}.

\begin{figure}[h]
\centering
\includegraphics[width=0.5\textwidth]{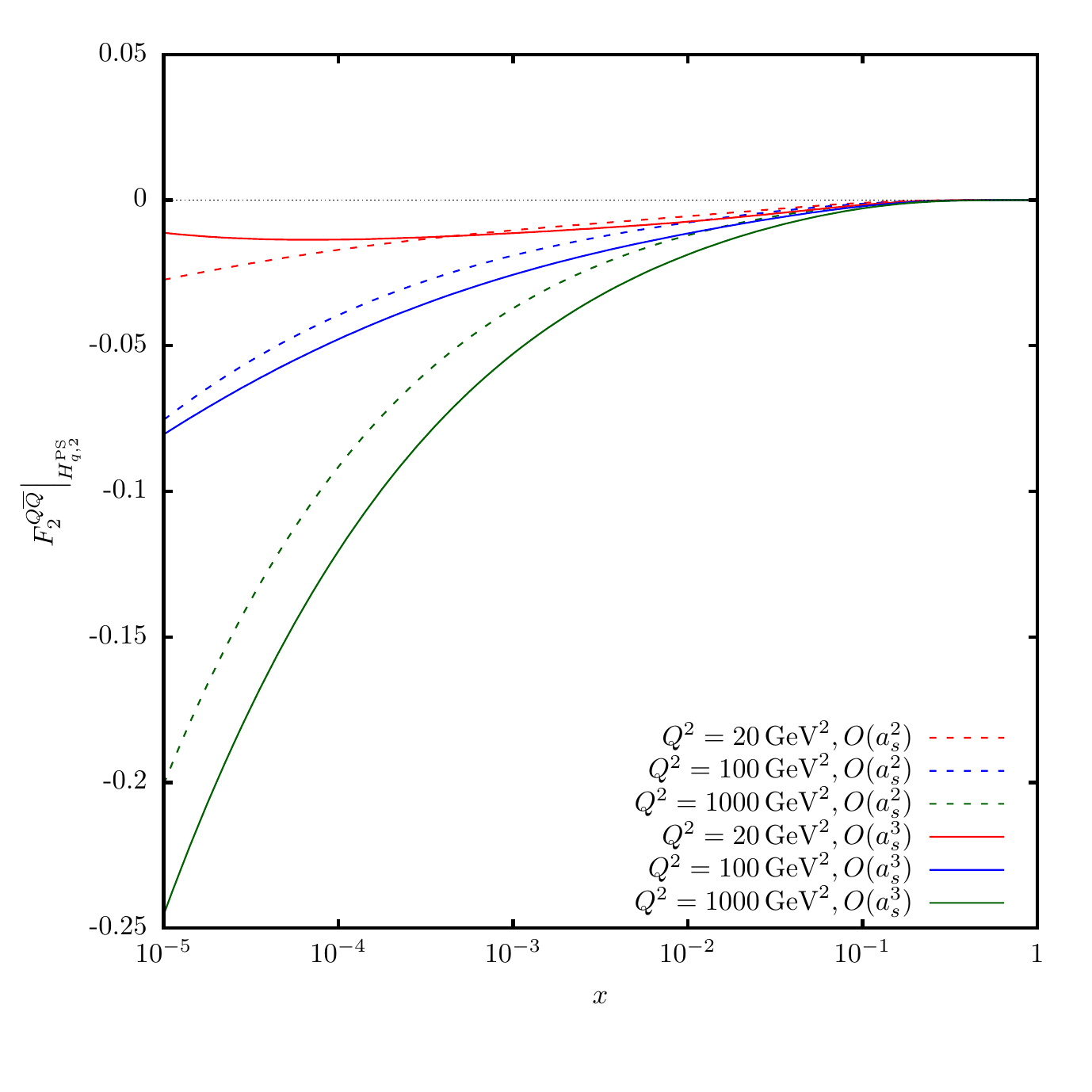}
\caption{The charm contribution by the Wilson coefficient $H_{Qq}^{PS}$ to the structure function $F_2(x,Q^2)$ as a 
function of $x$ and $Q^2$ choosing $Q^2 = \mu^2$ and $m_c = 1.59$\,GeV (on shell scheme) using the parton distributions
\cite{Alekhin:2013nda}; from \cite{Blumlein:2014zxa}.
\label{fig:jb}}
\end{figure}   

In the case of charged current interactions, including also single heavy quark excitation, the 1-loop calculation had to be repeated because of 
conflicting results in the literature. In Ref.~\cite{Blumlein:2011zu} we confirm the result obtained in \cite{Gluck:1996ve}. 
At next-to-leading order (NLO) we correct an earlier result given in \cite{Buza:1997mg} in Ref. \cite{Blumlein:2014fqa}. 

We have by now obtained also complete results for a series of operator matrix elements and massive Wilson coefficients at 3-loop order, 
respectively those of complete contributions to individual colour factors. The contributions of $O(N_F T_F^2 C_{F,A})$ to all OMEs
were calculated in Refs.~\cite{Ablinger:2010ty,Blumlein:2012vq} and those of $O(T_F^2 C_{F,A})$ to $A_{gg,Q}$ in \cite{Ablinger:2014uka}.
The solution of special higher topologies for ladder-, Benz- and $V$-graphs has been presented in Refs.~\cite{Ablinger:2012qm,Ablinger:2014yaa}.
So far complete 3-loop results have been obtained for the massive OMEs $A_{gq,Q}$ \cite{Ablinger:2014lka}, $A_{qq,Q}^{NS}$ \cite{Ablinger:2014vwa}
and $ A_{Qq}^{PS}$ \cite{Ablinger:2014nga}. In Figure~\ref{fig:jb} we illustrate the size of the 2- and 3-loop flavor pure singlet contributions, which
are significant.

In the course of these calculations, various mathematical and computer-algebraic technologies had to be developed, including new mathematical function 
spaces and analytic algebraic integration techniques. Recent reviews on those are given in Refs.~\cite{MATH}.

The present heavy flavour corrections will improve also the precision determination of the parton densities in the future, cf. e.g.~\cite{Alekhin:2013nda}.  
In particular, the planned high luminosity measurements at the EIC \cite{EIC,Boer:2011fh,Accardi:2012qut} or LHeC \cite{AbelleiraFernandez:2012cc} 
can also test these corrections.

%% file: Contributions/SvenMoch/moch.tex
\subsection{Precise parton distributions \\
  {\it S.~Moch}
}

The need for precise PDFs can best be illustrated with the current theoretical accuracy for 
predictions of the cross section for Higgs boson production in gluon--gluon fusion at the LHC.
The QCD radiative corrections are sizeable and require the computation of higher orders  
to achieve apparent convergence of the perturbative expansion and stability under scale variation.
Currently, approximate N$^3$LO (next-to-next-to-next-to-leading order) results
are available~\cite{deFlorian:2014vta,Anastasiou:2014lda} which display a
residual theoretical uncertainty of less than 5\%.
In the effective theory (limit of infinite top-quark mass) one obtains 
for a Higgs boson mass $m_H=125.4$\,GeV and a LHC centre-of-mass energy $\sqrt{S}=8$\,TeV 
at the scale $\mu_R=\mu_F=m_H/2$ close to the point of minimal sensitivity 
at approximate N$^3$LO 
\begin{equation}
  \label{eq:sn3lo}
  \sigma_{H}\bigl|_{\rm{N}^3\rm{LO}} 
  \,=\, 21.12\, \mbox{pb}\,\, 
  \,\pm\, 0.56\, \mbox{pb}\,\,\, (\mbox{th.~unc.})\,\,\,
  ^{+\, 0.29}_{-\,0.42}\, \mbox{pb}\,\,\, (\mbox{scale~unc.})
  \, ,
\end{equation}
where the theoretical uncertainty comes from estimating the exact N$^3$LO
corrections and the scale uncertainty is obtained from a simultaneous variation of
$\mu_R$ and $\mu_F$ by a factor of two around the central scale~\cite{deFlorian:2014vta}.
These two uncertainties are largely independent.

The dominant uncertainty in the prediction of eq.~(\ref{eq:sn3lo}) stems from
the PDF set and the value of the strong coupling constant $\alpha_s(M_Z)$ used,  
since the cross section is directly correlated with the gluon PDF and the value for $\alpha_s(M_Z)$.
In the case of eq.~(\ref{eq:sn3lo}), the PDF set of MSTW08~\cite{Martin:2009iq} 
has been used together with their nominal $\alpha_s(M_Z)=0.1171$ at NNLO.
However, such high $\alpha_s(M_Z)$ values have recently been challenged 
by high precision determinations yielding much lower values, see e.g.~\cite{Moch:2014tta}.
To illustrate the dependence on the choice of a PDF set, it suffices to consider 
cross section predictions at NNLO in QCD. 
For the MSTW08 PDF set one obtains 
\begin{equation}
  \label{eq:sn2lo-mstw}
  \sigma_{H}\bigl|_{\rm{NNLO}} 
  \,=\, 20.16\, \mbox{pb}\,\, 
  ^{+\, 0.22}_{-\,0.32}\, \mbox{pb}\,\,\, (\mbox{PDF})
  \, ,
\end{equation}
which is in contrast to the cross section calculated with the ABM12 PDF set~\cite{Alekhin:2013nda}
with a nominal $\alpha_s(M_Z)=0.1132(11)$ at NNLO,
\begin{equation}
  \label{eq:sn2lo-abm12}
  \sigma_{H}\bigl|_{\rm{NNLO}} 
  \,=\, 18.67\, \mbox{pb}\,\, 
  \,\pm\, 0.46\, \mbox{pb}\,\,\, (\mbox{PDF})
  \, .
\end{equation}
The deviation between eqs.~(\ref{eq:sn2lo-mstw}) and (\ref{eq:sn2lo-abm12}) 
is about 8\% and much larger than the residual theoretical uncertainty 
of less than 5\% given in eq.~(\ref{eq:sn3lo}) with the approximate N$^3$LO corrections.
It is at the level of $3-4\sigma$ standard deviations with respect to the 
uncertainty determined from the correlated experimental uncertainties in the fitted data.
Also the use of other global PDF fits, such as 
CT10~\cite{Gao:2013xoa}, HERAPDF (v1.5)~\cite{HERA1-comb,herapdfgrid:2011},
JR09~\cite{JimenezDelgado:2008hf} and the recent updates 
MMHT 2014\cite{Harland-Lang:2014zoa} and NNPDF (v3.0)\cite{Ball:2014uwa}, 
leads to a significant spread in the predicted Higgs cross sections, see, e.g. \cite{Alekhin:2010dd}. 

The reasons for the differences are, of course, known and well documented in the literature.
They originate from a number of sources which encompass specific choices in
the theoretical description of the data, the selection of data sets and the analysis procedures.
Examples for theory choices are the treatment of heavy quarks in deep-inelastic scattering (DIS)
as well as the renormalisation scheme for the heavy-quark masses~\cite{Alekhin:2010sv} 
(running masses for charm and bottom in ABM11~\cite{Alekhin:2012ig} and ABM12~\cite{Alekhin:2013nda}).
Keeping a fixed number of flavors \cite{Alekhin:2009ni} together with the
use of the approximate NNLO corrections for the Wilson coefficients of heavy-quark
electro-production~\cite{Bierenbaum:2009mv,Kawamura:2012cr} 
ensures the exact description of all terms up to $\alpha_s^3 \ln^k(Q^2/m^2)$ with $k=1, \dots, 3$.
Other examples are the use of target mass corrections in fits of lepton--proton scattering data~\cite{Alekhin:2012ig}
or the consistent extraction of power corrections (higher-twist contributions in DIS data at low scales)~\cite{Alekhin:2011ey}.
All these issues affect the gluon distribution and the fitted value of $\alpha_s(M_Z)$.
Regarding the selection of data sets a prominent example is the use of Tevatron jet data 
in a global fit at NNLO accuracy in order to constrain the gluon PDF.
The QCD description of hadro-production jet data is currently accurate only to
NLO so that missing large higher order corrections at NNLO~\cite{Ridder:2013mf} estimated to be of
the order of 30\% lead to a bias in the extracted gluon PDF~\cite{Alekhin:2012ig}.
Finally, details of the analysis procedure such as the proper account of error
correlations among data sets does affect the statistical interpretation~\cite{Alekhin:2012ce}.

In summary, the need for precision PDFs is driven by the experimental accuracy of cross section measurements. 
Checks of the specific theory assumptions made in PDF and $\alpha_s$ fits as well as 
checks of correlations between experimental data for different scattering processes 
at the LHC and their sensitivity to PDFs are urgently needed to consolidate 
our knowledge of the proton structure.

%% file: Contributions/HubertSpiesberger/spiesberger.tex
\subsection{Electroweak physics with HERA data \\
{\it H. Spiesberger}
}

With the LHC experiments, the focus of present research in particle 
physics has shifted towards the measurement of the Higgs boson 
properties and the search for new physics. A conclusive 
interpretation of future data will still rely on a precise 
knowledge of the Standard Model parameters. In particular, 
the weak mixing angle will be important, should new phenomena be 
observed. At present its value is dominated by two high-precision 
measurements, which disagree, however, by almost $3\sigma$. Any 
additional information is important to narrow down possible new 
physics scenarios.

There have been various analyses, both by H1 and ZEUS, of 
inclusive data from $e^-$ and $e^+$ scattering, also including 
polarised beams, to determine electroweak coupling parameters, 
but none has used the full data set in a combined fit. 
For example, H1 has performed a fit to the Standard Model 
prediction based on 120\,pb$^{-1}$ of unpolarized $e^{\pm}$ 
data and obtained a 1.8\,\% determination of the weak 
mixing angle, $\sin^2 \theta_W = 0.2151 \pm 0.0040$ (with 
an additional $^{+0.9}_{-0.5}$\,\% theoretical uncertainty) 
\cite{Aktas:2005iv}. The complete data sets of H1 and ZEUS 
correspond to almost 1000\,pb$^{-1}$, thus one should expect 
a much better precision if these data sets are combined. 

\begin{figure}[h]
\begin{center}
\includegraphics[trim={0cm 0cm 14cm 22cm},clip,width=0.6\textwidth]{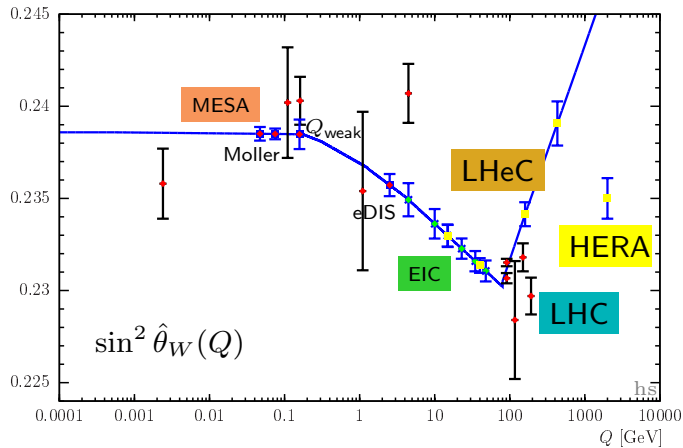}
\caption{
\footnotesize{
Various existing (red points with black error bars \cite{Agashe:2014kda}) 
and possible future measurements of the running ($\overline{\rm MS}$) 
weak mixing angle, including a potential result from a SM fit to HERA 
data: $\Delta \sin^2 \theta_W = \pm 0.0011$, corresponding to $\Delta 
M_W = 55$\,MeV.
}
}
\label{Fig:Sec:electroweak}
\end{center}
\end{figure} 

The sensitivity of a combination of neutral and charged 
current data with polarized electrons had been estimated 
some 20 years ago \cite{Beyer:1995pw} based on $10^3$\,pb$^{-1}$ 
and assuming a 1\,\% systematic error. A SM fit in the 
on-shell scheme combined with a very conservative expected 
precision for the top-quark mass of $\Delta m_t = \pm 5$\,GeV 
would provide a $W$-mass determination with a precision 
$\Delta m_W = \pm 55$\,MeV, corresponding to $\Delta \sin^2 
\theta_W = \pm 0.0011$. Uncertainties in the PDFs have 
now reached the level of a percent and would therefore 
indeed not obviate such a precision. It remains to be seen 
by how much the uncertainty on $m_W$ is affected by (i) the 
different mixture of electron and positron data with about 
the same total amount of luminosity as has been assumed in 
\cite{Beyer:1995pw}, and (ii) the different degree of polarisation. 
Such a 0.5\,\% measurement of $\sin^2 \theta_W$ would not compete 
in precision with the available high-precision results, but is 
complementary and therefore a useful addition to the global 
picture of electroweak precision measurements (Fig.\ 
\ref{Fig:Sec:electroweak}). 

HERA offers the unique possibility to measure the weak vector and 
axial--vector coupling constants of quarks. The four independent 
measurements with polarised beams, $e_L^-$, $e_R^-$, $e_L^+$, 
$e_R^+$ carry enough information to separately determine $v_f$ 
and $a_f$ for up- and down-type flavours. Preliminary results 
exist \cite{Aktas:2005iv,Rizvi:2009zz}, but from previous studies 
\cite{Cashmore:1996je} one can conclude that there is a considerable 
potential for improving the precision of these measurements. 


%% file: Contributions/KatarzynaWichmann/wichmann.tex
\subsection{Future electroweak and contact interaction fits\\
{\it K. Wichmann, V. Myronenko, O.Turkot, A.F. Zarnecki}
}

\subsubsection{HERA polarised data}

The H1 and ZEUS collaborations published their final results for 
inclusive DIS at high photon virtuality $Q^2$ with 
longitudinally polarised lepton beams at HERA~\cite{HERA1-comb,zfinal,H1final,dis}. 
The data were taken both for electron and positron incoming beams and for 
right-handed and left-handed polarisation of the lepton beam. 
This make these measurements perfect to study processes
sensitive to the beam polarisation.

\subsubsection{Combined electroweak and QCD global fits}

Inclusive DIS cross sections measured at HERA are essential for studying 
electroweak physics and performing a consistent QCD analysis to reveal the proton 
structure and obtain PDFs. 
Such studies were performed previously by H1 and ZEUS using data from HERA\,I 
period (no lepton beam polarisation). 
Using data from the HERA\,II period with polarised leptons gives additional sensitivity to 
the measurements of light quark couplings to the $Z^0$ boson, in particular to vector couplings. 
It is also possible to measure the weak mixing angle and the mass of the $W$ boson. 
Exploiting complete data samples from H1 and ZEUS simultaneously and taking into account 
correlated systematic uncertainties and global normalisations  gives 
a significant improvement in precision.
The study of the electroweak parameters can be done either using HERA only data 
or with additional constraints from other  experiments. 
Such studies can be performed within the HERAFitter framework~\cite{herafitter} to coherently take
into account the PDF uncertainties and the correlated uncertainties of 
inclusive measurements.

\subsubsection{Contact interactions}

In a search for contact interactions (CI) some channels are sensitive to the 
lepton polarisation.
In previous CI studies, the PDF uncertainty was a dominant source of 
systematic uncertainties and 
it was used to vary model predictions in the limit setting procedure.
The fact that PDFs used in the analysis 
already include HERA high-$Q^2$ data was neglected.
This was acceptable for the HERA\,I data, dominated by the statistical uncertainty,  
but for the full HERA data it has to be taken properly into account.

A procedure where the same high-$Q^2$ DIS data are first included in the PDF fits 
used \mbox{to describe} the data (assuming validity of the SM) 
and then for the CI mass scale limit calculation (testing possible contributions beyond SM) 
is  incorrect.
A possible BSM contribution, modifying electron--quark interactions at the highest 
$Q^2$, would also be reflected in the PDF fit. 
Resulting PDFs \mbox{would be} biased and the SM predictions 
would also include some BSM contributions ``hidden'' in the PDFs. 
As a result, one could still obtain a good agreement of the data with the biased SM 
predictions and the limits on the BSM parameters calculated with these PDFs could be 
artificially overestimated.

The only proper procedure to set limits on the BSM models
using the data included in the PDF fits is to perform a combined analysis, i.e. include 
possible contribution from the BSM processes in the QCD fit to the data. 
The resulting limits on the
model parameters might not be competitive with the expected LHC limits but
these would be the first limits set in the strictly correct way.

%% file: jets.tex
\section{Jets and hadronic final states}
\label{sec:jets}

\input{Contributions/IainStewart/stewart}

\input{Contributions/UriKarshon/karshon}

\input{Contributions/ErichLohrmann/lohrmann}

%% file: Contributions/IainStewart/stewart.tex
\subsection{Precision jet physics in deep inelastic scattering \\
{\it D. Kang, C. Lee, I.W. Stewart}
}

Event shapes provide a key method of measuring jets in DIS. This was done successfully by H1 and ZEUS~\cite{Adloff:1997gq,Adloff:1999gn,Aktas:2005tz,Breitweg:1997ug,Chekanov:2002xk,Chekanov:2006hv} and compared with theoretical calculations with next-to-leading-logarithmic (NLL) resummation~\cite{Antonelli:1999kx,Dasgupta:2002dc}.  Here we consider
the event shape, DIS thrust, $\tau$, which is defined in the Breit frame using the momentum of the exchanged $\gamma$ or $Z$ boson to determine the $z$-axis, $q=(0,0,0,Q)$. It can be measured solely from events in the current hemisphere where $z>0$ via
$\tau = 1 - (2/Q) \sum_{i\in {\cal H}_J} p_{i z}$, thus avoiding the lack of detector coverage in parts of the beam region. The event shape $\tau$ also does not suffer from non-global logarithms~\cite{Dasgupta:2002dc}. 

Recently an all orders factorisation theorem was derived for $d\sigma/d\tau$~\cite{Kang:2013nha}, which enables higher order perturbative results to be obtained, and a more rigorous treatment of power corrections,
\begin{align} \label{factthm}
   \frac{1}{\sigma_0} \frac{d\sigma}{dx dQ^2 d\tau}
    = Q\, H(Q,x,\mu)\!\! \int\!\! dt_B dt_J d^2p_\perp\, B_q(t_B,x,\vec p_\perp^{\,2},\mu)\: J_q(t_J-\vec p_\perp^{\,2},\mu)\:
   S\Big( Q\tau -\frac{t_B+t_J}{Q},\mu\Big).
\end{align} 
Results with a resummation of the singular $\alpha_s^k\ln^j\tau/\tau$ terms at next-to-next-to-leading-log (NNLL) order were given in~\cite{Kang:2013nha}. 
Here we extend this analysis to one higher order, N$^3$LL, by exploiting the recent 2-loop calculation of the quark beam function $B_q$~\cite{Gaunt:2014xga,Gaunt:2014xxa}, the 2-loop DIS soft function $S$~\cite{Kang:2015new,Kelley:2011ng,Monni:2011gb}, and known results for the 2-loop hard function $H$ and jet function $J_q$, and their 3-loop anomalous dimensions~\cite{Matsuura:1988sm,Becher:2006qw,Moch:2004pa}. The smaller nonsingular contributions to $d\sigma/d\tau$ are also now known analytically at ${\cal O}(\alpha_s)$~\cite{Kang:2014qba}, while numerical results are available at ${\cal O}(\alpha_s^2)$~\cite{Catani:1996vz,Graudenz:1997gv}. Power corrections are encoded by a hadronic matrix element $\Omega_1$ appearing in $S$, using formalism developed in Refs.~\cite{Lee:2006fn,Lee:2006nr,Hoang:2007vb,Mateu:2012nk,Kang:2013nha}.  
(The DIS thrust $\tau$ is equal to the Breit frame 1-jettiness $\tau_1^b$, and hence belongs to the class of 1-jettiness event shapes~\cite{Stewart:2010tn}. Results for other 1-jettiness DIS variables were obtained in Refs.~\cite{Kang:2012zr,Kang:2013lga,Kang:2013wca,Kang:2013nha}, currently up to NNLL order.)

Fits for $\alpha_s(m_Z)$ in the tail region of the DIS $\tau$ distribution, should simultaneously fit for the power correction $\Omega_1$ (similar to the highly successful fits for the $e^+e^-$ thrust event shape in~\cite{Abbate:2010xh}). This is facilitated by  considering $d\sigma/d\tau$ from multiple $x$ and $Q$ values. Interestingly, the factorisation theorem in Eq.~(\ref{factthm}) remains valid for relatively small $x$, and  the fractional contribution from the nonsingular corrections even decreases with decreasing $x$, as shown at ${\cal O}(\alpha_s)$ in~\cite{HERAtalk:2014}.

\begin{figure}[t!]
\includegraphics[width=0.49\columnwidth]{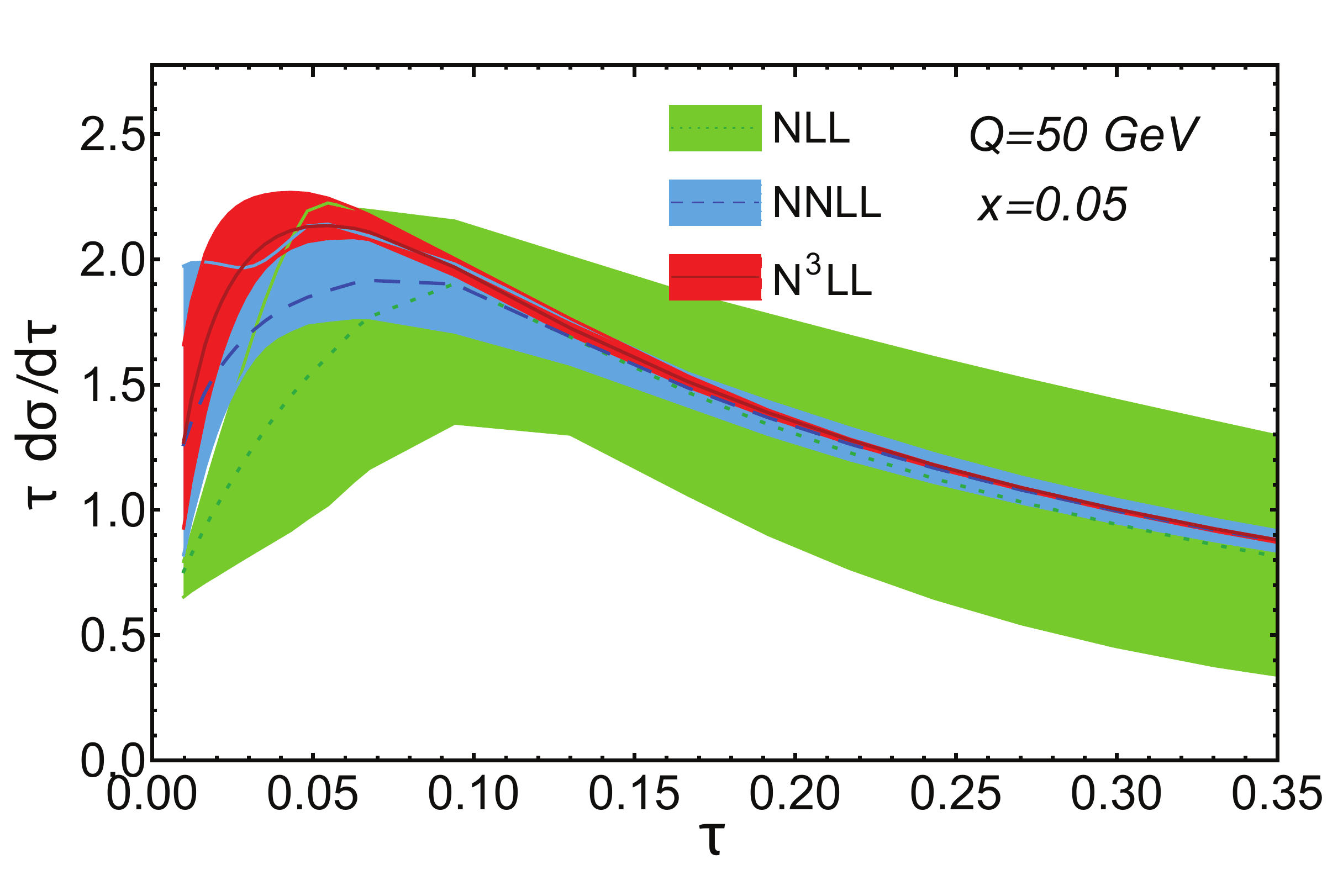} \hspace{0.1cm}
\includegraphics[width=0.49\columnwidth]{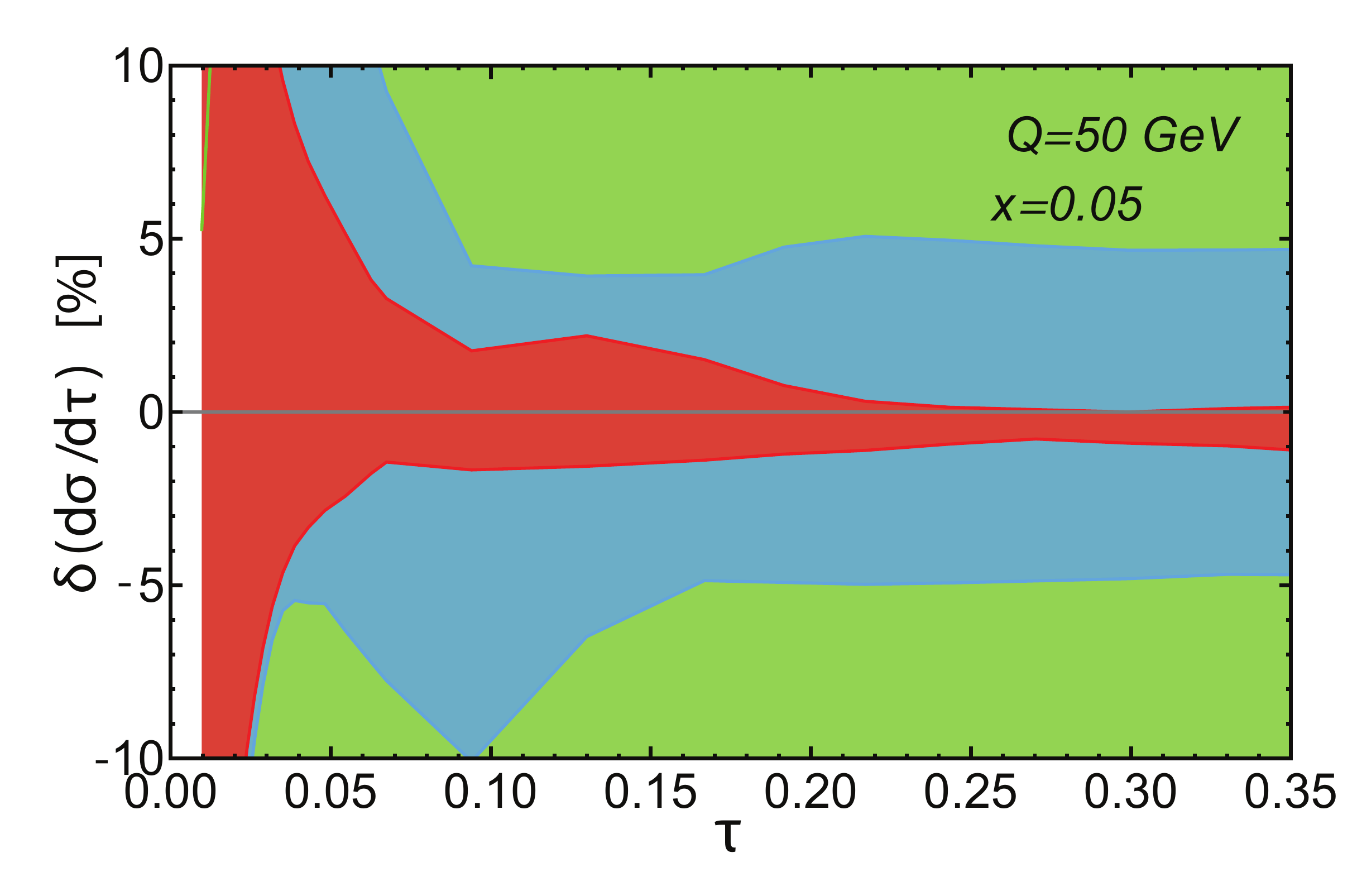} 
\caption{\setlength\baselineskip{13pt} Convergence of the DIS thrust distribution. Results at three orders are shown along with their perturbative uncertainty. Left panel shows $\tau d\sigma/d\tau$. Right panel shows the relative uncertainty for $d\sigma/d\tau$.}
\label{fig:sigma_conv}
\end{figure}

\begin{figure}[t!]
\begin{center}
\includegraphics[width=0.7\columnwidth]{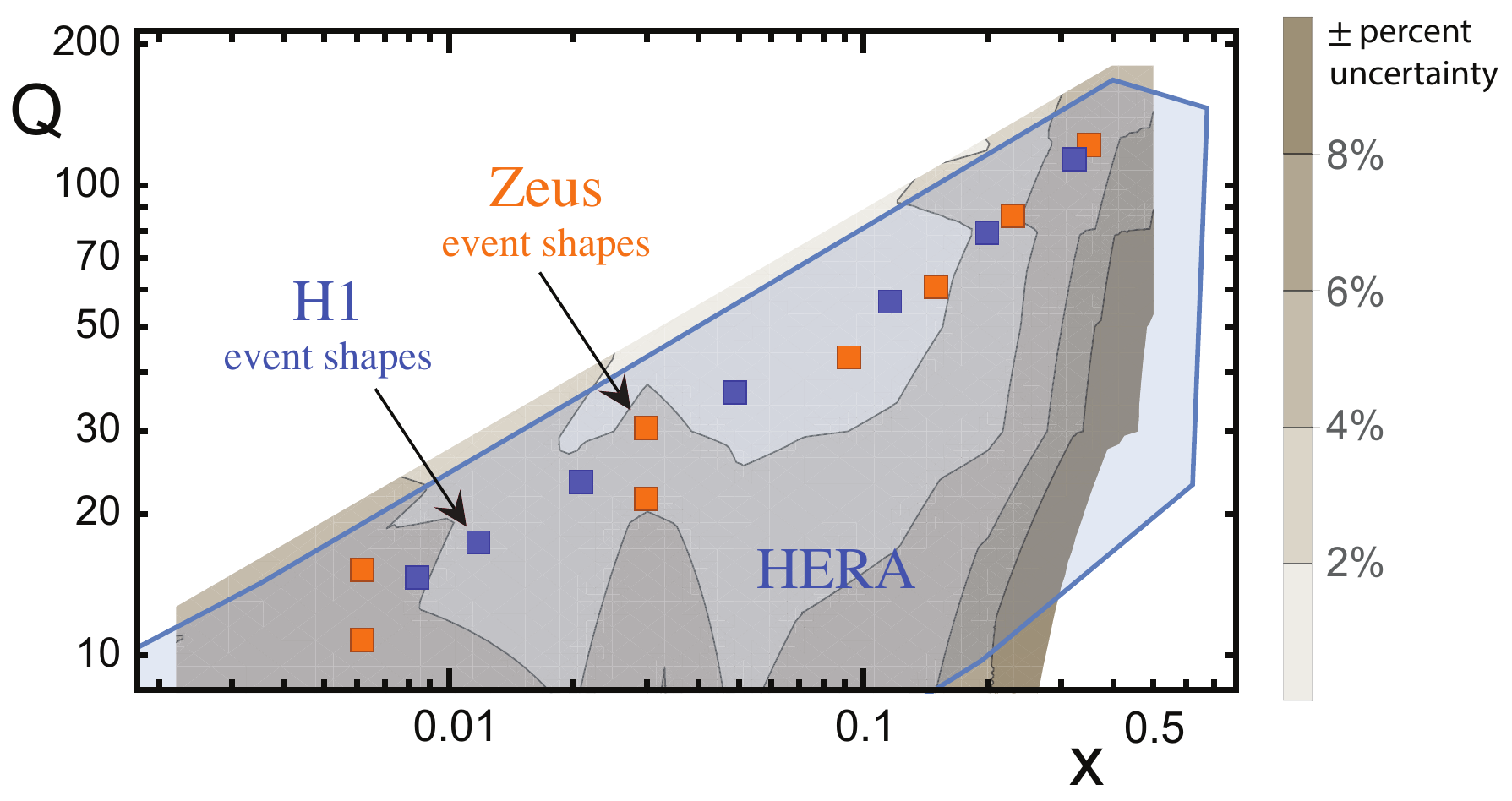}
\caption{ \setlength\baselineskip{13pt} Percent uncertainty of our N$^3$LL cross section for the region in $x$ and $Q$ accessible at HERA. Uncertainties are for the tail of the DIS thrust distribution which can be used to measure $\alpha_s(m_Z)$. Also shown are the points used for past DIS event shape measurements.}
\label{fig:HERA_region}
\end{center}
\end{figure}

\begin{figure}[t!]
\includegraphics[width=0.49\columnwidth]{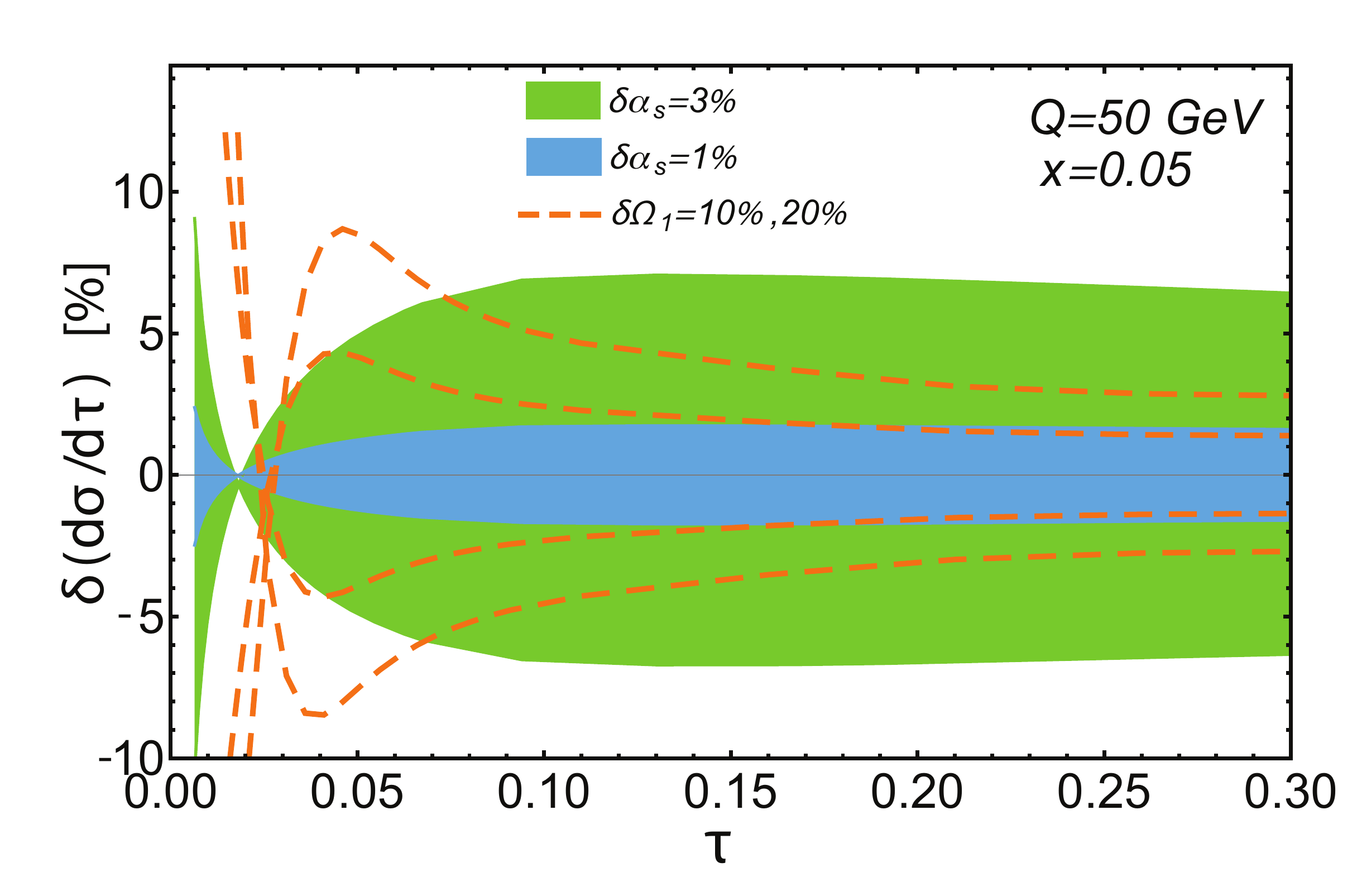}
\hspace{0.1cm}
\includegraphics[width=0.49\columnwidth]{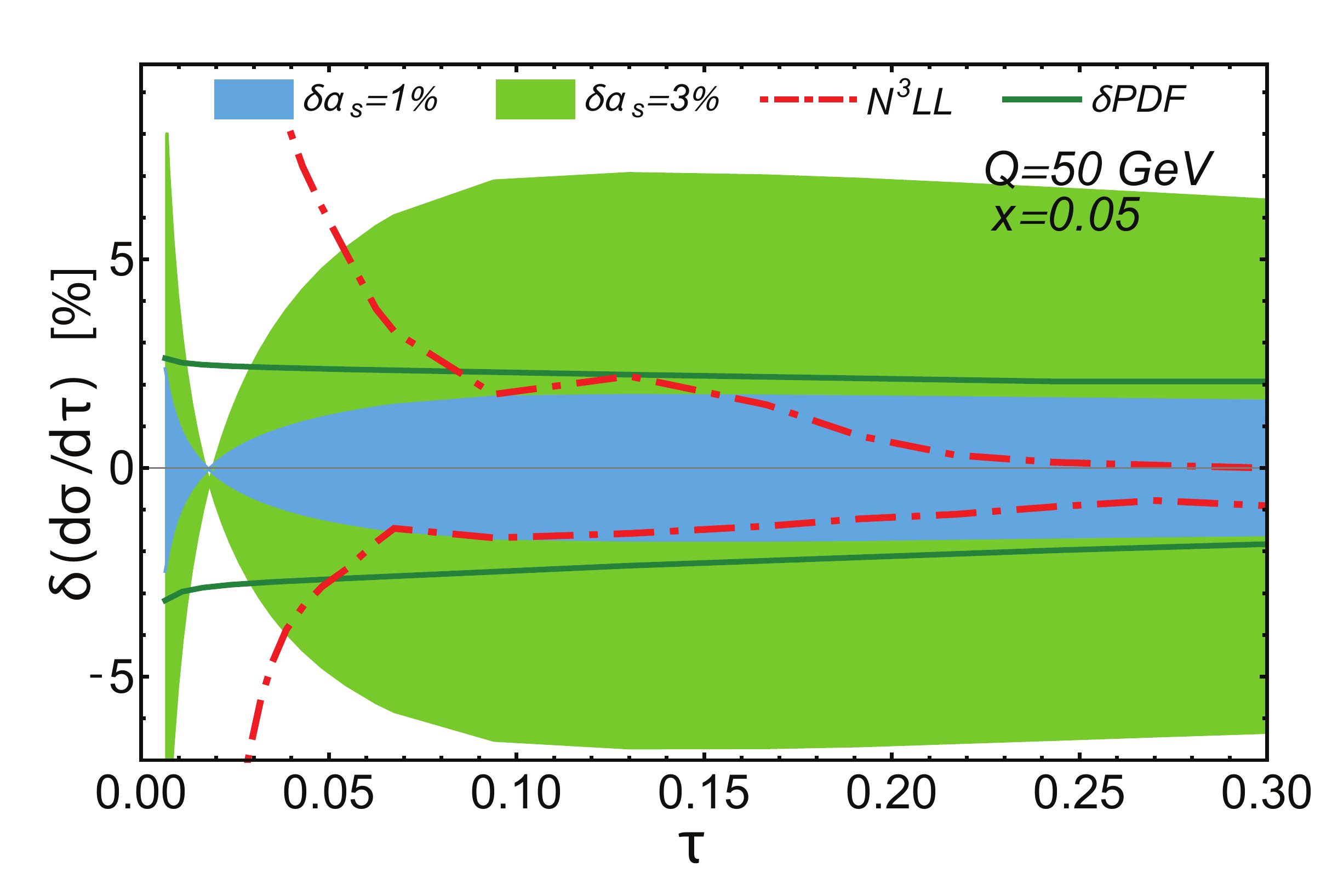}
\caption{ \setlength\baselineskip{13pt} Sensitivity of the DIS thrust cross section to changes in $\alpha_s(m_Z)$ and $\Omega_1$ (left panel) and compared with PDF and N$^3$LL uncertainties (right panel).   }
\label{fig:alphas}
\end{figure}

In Fig.~\ref{fig:sigma_conv} we show the convergence of the DIS thrust cross section and decrease in the perturbative resummation uncertainty when going from NLL to NNLL to N$^3$LL order. Results are displayed for a representative value of $x$ and $Q$, while cross sections for other values can be found in~\cite{HERAtalk:2014}. In Fig.~\ref{fig:HERA_region} we show the  percent uncertainty of $d\sigma/d\tau$ for various values of $x$ and $Q$ in the region accessible by HERA, demonstrating that the theoretical resummation uncertainties become as low as 2\% in accessible regions of the phase space. Values are obtained as the average uncertainty in $d\sigma/d\tau$ in the tail region $0.15<\tau<0.35$. In Fig.~\ref{fig:alphas} we show how much the cross section changes with variations of the input parameters $\alpha_s(m_Z)$ and $\Omega_1$, as well as comparing the $\alpha_s(m_Z)$ sensitivity to the N$^3$LL resummation uncertainties, and to the uncertainties from the NNLO  MSTW parton distributions~\cite{Martin:2009iq}. Figures for other values of $x$ and $Q$ are available in~\cite{HERAtalk:2014}.
The degeneracy between $\alpha_s(m_Z)$ and $\Omega_1$ is broken by measurements at multiple $Q$. The theoretical precision of our N$^3$LL cross section indicates that measurements with 1--2\% uncertainty in $\alpha_s(m_Z)$ should now be possible. A measurement of $\Omega_1$ from DIS is also of broader use, since this same $\Omega_1$ parameter occurs in $pp\to Z+1$-jet, where it yields the power correction for the jet mass that is linear in the jet radius~\cite{Stewart:2014nna}.

%% file: Contributions/UriKarshon/karshon.tex
\subsection{Why and how to search for charm pentaquarks with the HERA data \\
{\it U. Karshon}
}
                                                                             
If a pentaquark is found and confirmed, it will provide a major
   testing ground for low-energy QCD dynamics. 
Many theorists think that heavy pentaquarks are more likely to exist 
 than strange ones~\cite{Marek}.
Some of them               predict a       stable heavy charm pentaquark 
with $M < 2.8$~GeV below the $DN$ threshold~\cite{Jaffe}. The
negative experimental results~\cite{cpq-neg} for a $u u d d \bar c$ charm pentaquark 
 $\Theta^0_c\to D^* N, D N$ at $3.1$~GeV, after its initial observation~\cite{H1}, could be due to
the stability of charm pentaquarks.                                             
            Thus they should be searched for in weak decays. 
Both the H1 and ZEUS experiments had a microvertex detector (MVD) enabling reconstruction of 
secondary vertices from weakly decaying particles.
Weakly decaying charm-strange pentaquarks could also exist~\cite{Zvi},      
                       e.g.  {$P^0_{\bar c s}=\bar c s u u d$ and 
      $P^-_{\bar c s}=\bar c s d d u$}.                         
 
   If  {$\Theta^0_c$} exists with a mass below $\approx 2.8$~GeV,
                                it will decay weakly, e.g.    
{$\Theta^0_c\to K^0_s\pi^- p$}                          
        where $p,\pi^-$ come from a secondary MVD vertex    
                       and   $K^0_s$   
                             points to this vertex   
or {$\Theta^0_c\to K^+\pi^-\pi^- p$}  where all 4 particles come from a secondary vertex.
 
   If  $P_{\bar c s}$  exists below the strong decay
   threshold, it will decay weakly, e.g.:
 
$$
P^0_{\bar c s}\to\phi\pi^- p,~~ \phi\to K^+ K^-;
P^0_{\bar c s}\to K^{*0}K^- p,~~ K^{*0}\to K^+\pi^-;
$$
$$
P^0_{\bar c s}\to (K\bar K)^0 p\pi^-;
P^0_{\bar c s}\to \Lambda K^+\pi^-,
$$
$$
P^-_{\bar c s}\to\phi\pi^-\pi^- p,~~\phi\to K^+ K^-;
P^-_{\bar c s}\to K^{*0}K^-\pi^- p,~~K^{*0}\to K^+\pi^-;
$$
$$
P^-_{\bar c s}\to\Lambda\pi^-\pi^- K^+;~~~~
P^-_{\bar c s}\to\Lambda\pi^- K^0      \ \     {\rm (a \ very \ clean \ channel!)}
$$
where all decaying particles come from a secondary MVD vertex.  
Any detection of      {$p (\bar p)$} or      {$\Lambda (\bar \Lambda)$} emerging from a MVD secondary
vertex (unless coming from a $\Lambda_c$) could hint at the
existence of an exotic baryon~\cite{MaZvi}.

%% file: Contributions/ErichLohrmann/lohrmann.tex
\subsection{A new search for instantons \\
{\it E. Lohrmann}
}

The existence of instantons is predicted by the Standard Model, but
so far they have not been observed. Their discovery would be a
confirmation of an important prediction connected with the QCD vacuum. 

Ringwald and Schrempp have identified kinematic regions in
$ep$ DIS, which allow an approximate calculaton
of instanton-induced processes~\cite{1}. A Monte Carlo generator
QCDINS~\cite{2} allows simulation of DIS instanton events in detail.  

A search for instanton events has been carried out by the H1~\cite{3}
and ZEUS collaborations~\cite{4}. In these searches DIS
events are selected and cuts designed to enhance the instanton contribution
of the sample are applied.

In the meantime, the conditions for an instanton search have drastically
improved. The work of ref.~\cite{4} was carried out with an integrated
luminosity of 38\,pb$^{-1}$. Now, about 10 times this luminosity
is available. Moreover, A.N. Barakbaev and E. Boos have proposed a new variable~\cite{5},
which allows a much better discrimination of instanton and normal DIS
events. It is based on the large difference of the transverse 
momentum/eta correlation between instanton and normal events. 

 In ref.~\cite{4} 
a conservative estimate of an upper limit for the fraction of instanton 
events in the sample was made, which is independent of the Monte Carlo 
background estimate of normal DIS events. This estimate is given
by $\sigma_d$/$\sigma_{th}$, where $\sigma_d$ is the number of data
events after cuts and $\sigma_{th}$ is the number of instanton 
events predicted by theory after cuts. 

Starting from the DIS events sample of ref.~\cite{4},
corresponding to an integrated luminosity of 38\,pb$^{-1}$, one
obtains values for $\sigma_d$/$\sigma_{th}$, according to the
new algorithm of ref.~\cite{5}, as shown in the table (data in the $\gamma$--proton
CM system are used). For comparison,
corresponding values for the analysis published in ref.~\cite{4}
 are also shown.
The comparison shows the advantage gained by the new algorithm.
It now formally excludes the theoretically predicted cross section,
but probably is still within the uncertainties of the theory.

\begin{table}[h]%
\centering
\caption{Ratio $\sigma_d$/$\sigma_{th}$, the ratio of signal events to 
events predicted by instanton theory, for an integrated luminosity
of 38\,pb$^{-1}$.}
\begin{tabular}{|r|r|r|}
\hline
  ZEUS ref.~\cite{4}  & cut      &  $\sigma_d$/$\sigma_{th}$    \\
              & 16.4    & 4.5 $\pm$ 0.2 \\
              & 10.1    & 3.1 $\pm$ 0.2 \\
              & 5.5     & 2.4 $\pm$ 0.3  \\
              & 2.7     & 2.1 $\pm$ 0.5   \\
\hline
 algorithm ref.~\cite{5} & cut     & $\sigma_d$/$\sigma_{th}$  \\
              & 0.45    & 6.7 $\pm$ 0.2  \\
              & 0.35    & 1.7 $\pm$ 0.2  \\
              & 0.30    & 0.9 $\pm$ 0.2  \\
              & 0.25    & 0.22 $^{+0.29}_{-0.13}$ \\
\hline
\end{tabular}
\end{table}  
The H1 Collaboration has conducted a new search~\cite{instantons} for
instantons, using the full HERA statistics. In a preliminary
result they set an upper limit on the instanton cross section
of a factor about 6 below the theoretically predicted value. However,
this value has some uncertainties, like e.g. the background subtraction 
from the DIS Monte Carlo calculation. Using the new algorithm could probably
further improve their result.

%% file: common.tex
\section{Physics topics common with other experiments}
\label{sec:common}

\input{Contributions/RonanMcNulty/mcnulty}

%% file: Contributions/RonanMcNulty/mcnulty.tex
\subsection{Common physics between HERA and LHCb\\
{\it R. McNulty}
}

\subsubsection{PDFs at HERA and LHCb}

The effort to understand the proton structure extends from HERA to the LHC,
both in developing a coherent picture at different parton fractional momenta, $x$,
and energy scales, $Q^2$, and in making predictions for proton--proton interaction rates.
The kinematic region accessible at the LHCb experiment, 
which is instrumented between pseudorapidities of 2 and 5,
is of particular interest as it
involves collisions between one high-$x$ parton, already well measured  at HERA, and one low-$x$ parton, which is either unknown or requires a large
DGLAP extrapolation in $Q^2$ of HERA results.

\begin{figure}[htb]
\centerline{\includegraphics[height=7cm]{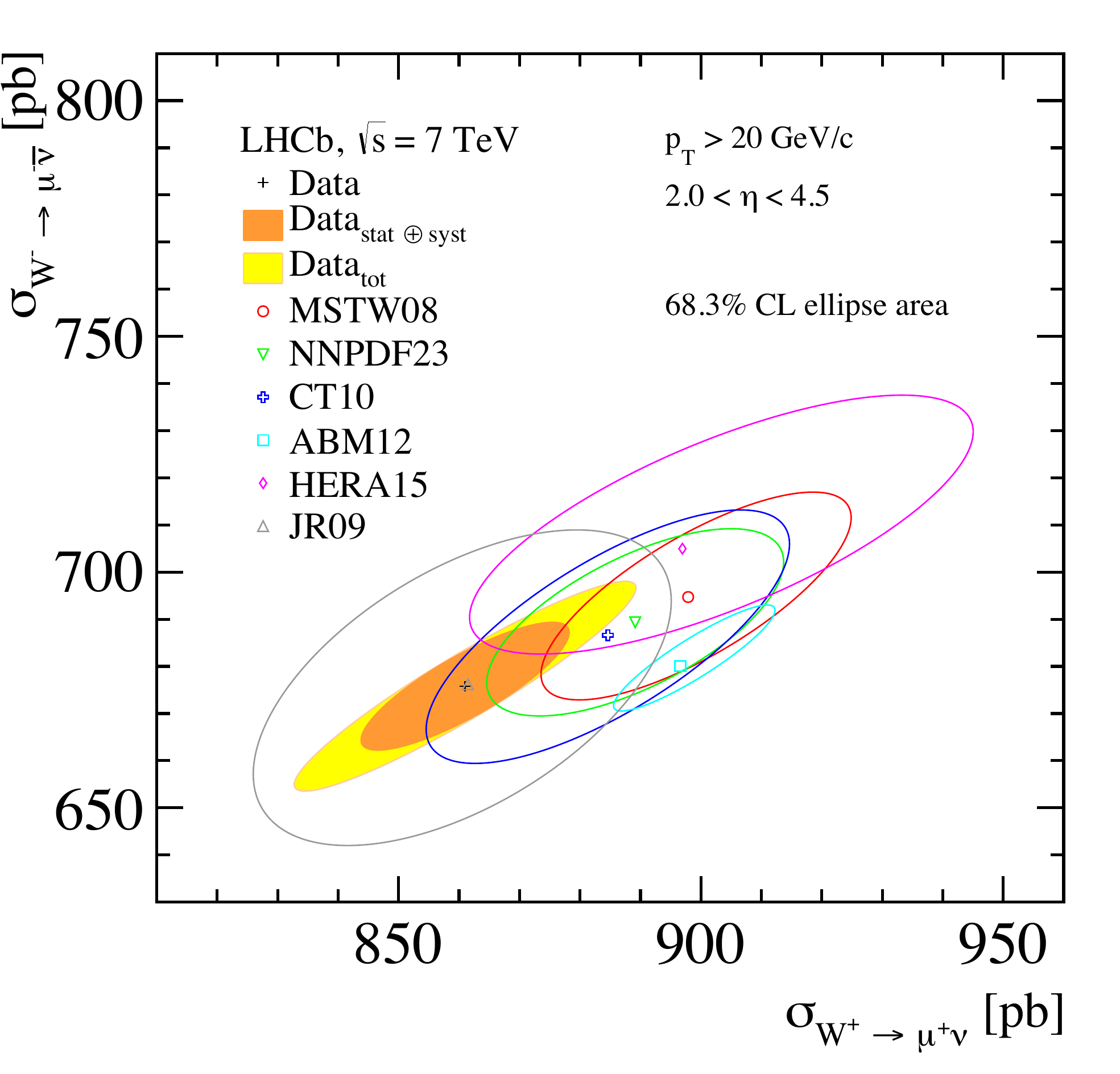}}
\caption{
Cross sections for  $W^+$ and $W^-$ bosons production inside the LHCb
kinematic region compared to NNLO predictions obtained with various PDF sets.
\label{fig:wz}}
\end{figure}

LHCb has made differential cross-section measurements~\cite{Aaij:2012vn,Aaij:2014wba} 
of $W$ and $Z$ boson
production at $\sqrt{s}=7$ TeV, which both test the Standard Model and constrain the PDFs.  
The ratio of $W^+$ to $W^-$ boson production provides strong constraints on the $u/d$
quark distributions, and the experimental ratio has been measured with a precision of better than 1\%
using data corresponding to an integrated luminosity of about 1 fb$^{-1}$ (see Fig.~\ref{fig:wz}).

\subsubsection{Diffraction at HERA and LHCb}

Photoproduction of vector mesons at HERA and the LHC proceeds through the interaction
of a photon and a Pomeron (a colour neutral combination of two gluons).  At HERA the photon
couples to the electron; at the LHC its source is the proton, leading to an unusual 
experimental topology for a hadron collider of a single meson and no additional particles in the
final state.

\begin{figure}[ht]
\centerline{\includegraphics[height=7cm]{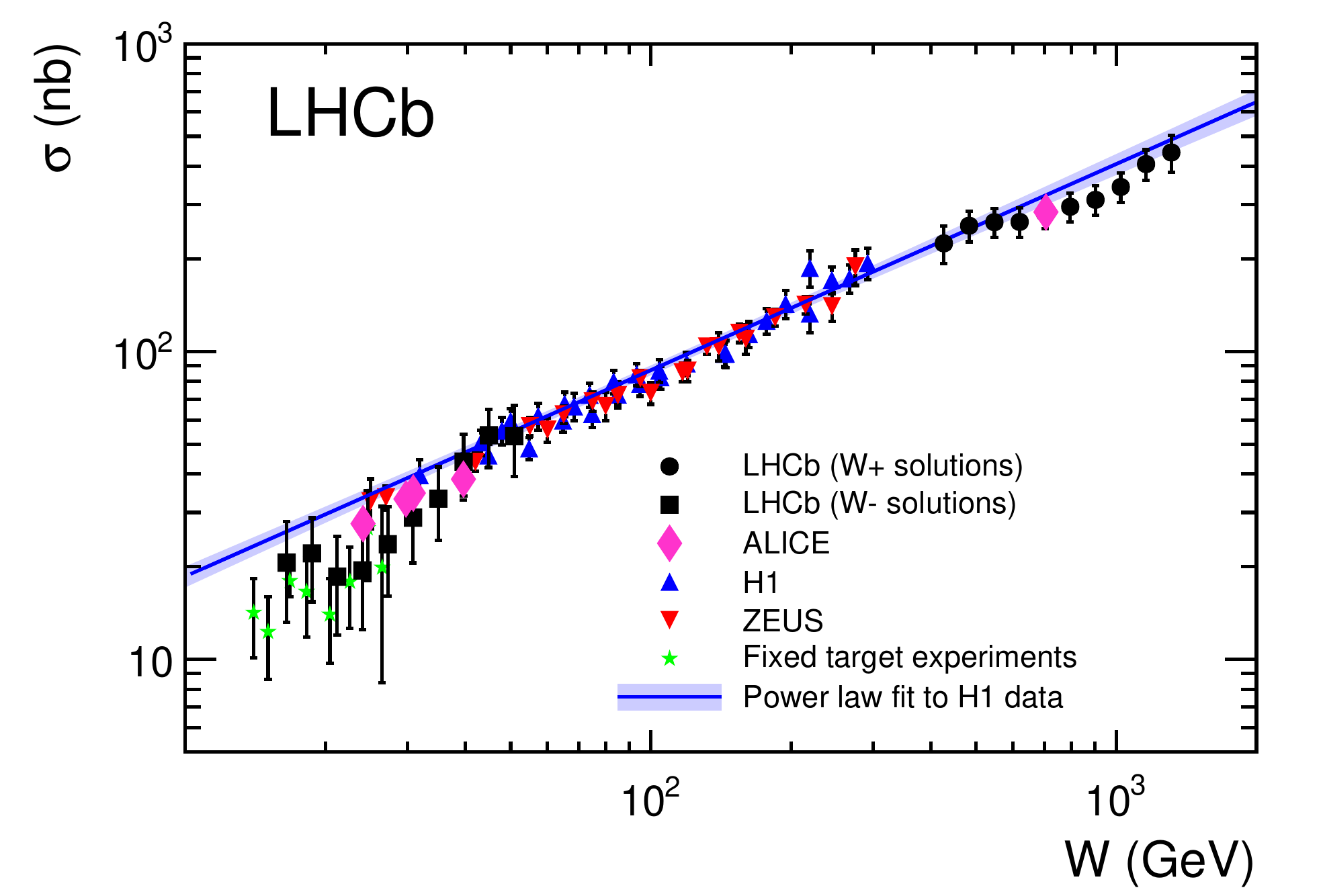}}
\caption{
Cross section for $J/\psi$ photoproduction as a function of the photon--proton centre-of-mass energy 
for various experiments.
\label{fig:jpsi}}

\end{figure}

LHCb has measured the central exclusive production of 
single $J/\psi$ mesons at $\sqrt{s}=7$ TeV~\cite{Aaij:2014iea} and the results can
be compared to HERA  photoproduction results, with some mild model dependence to
account for the photon flux, the gap survival probability, and a two-fold ambiguity in
knowing which of the two protons the photon radiated from.
The comparison is shown in Fig.~\ref{fig:jpsi}, which also plots the results of
fixed target experiments and of proton--lead collisions measured by the 
ALICE collaboration~\cite{TheALICE:2014dwa} at the LHC.
A consistent picture emerges of $J/\psi$ photoproduction across a wide range of $W$,
the centre-of-mass of the photon--proton system.
This is a promising channel to access the low-$x$ gluon PDF~\cite{Jones:2013pga},
where in the upcoming $\sqrt{s}=13$ TeV running, LHCb will have sensitivity
down to $x\approx2\times 10^{-6}$. 

These results are also important in our understanding of diffraction.  
The extensive programme of work at HERA now continues at the LHC where Pomeron--Pomeron
collisions can be observed, as for example in the recent observation of the central
exclusive production of double $J/\psi$ mesons.~\cite{Aaij:2014rms}.

%% file: diffraction.tex
\section{Diffraction and low-$x$ physics}
\label{sec:diffraction}

\input{Contributions/MartaRuspa/marta}

\input{Contributions/LeszekMotyka/motyka}

\input{Contributions/MichelSauter/sauter}

%% file: Contributions/MartaRuspa/marta.tex
\subsection{Future prospects for diffraction at HERA\\
{\it M. Ruspa}
}

By looking at the HERA harvest in diffraction with the aim to identify if and where the data can still tell something, what emerges are on the one hand open and critical issues, on the other hand the potential of measurements that were never carried out but are feasible with the latest high-statistics samples. There are also measurements that, though already performed, would benefit from the latest data. An overview is presented in the following. 

\vspace{0.3cm}
The {\bf critical issues} are related to the QCD factorisation theorem and the portability of HERA diffractive data to hadron--hadron environments: 

\noindent
- {\it Diffractive Parton Distribution Functions.} H1 and ZEUS published during the years a wealth of data on the diffractive structure function $F_2^D$~\cite{Chekanov20091,Chekanov:2004hy,Aktas:2006hx,Aaron:2010aa,Aaron:2012ad,Aktas:2006hy,Chekanov:2005vv,Chekanov:2008cw}. As in inclusive DIS, these data can be used as input to NLO DGLAP fits. A few sets of diffractive PDFs (DPDFs) have been extracted over the years by H1 and ZEUS~\cite{Chekanov:2009aa,Aktas:2007bv,Aktas:2006hy}. There are differences among the various fits, notably for the gluon density, which reflects the uncertainty in this quantity. 
A consistent determination of the DPDFs, felt as mandatory by the diffractive community outside HERA, can come only from the HERA experiments 
as only the HERA collaborations can understand the differences between the data. Adding diffractive dijet data in the fits has already proven to provide extra constraining power to the gluon density~\cite{Chekanov:2009aa,Aktas:2007bv}. Very recent high precision H1 dijet data~\cite{Andreev:2014yra} would thus provide a valuable further input. 

\noindent
- {\it Mechanism of factorisation breaking.}
A key physics issue is whether the DPDFs are or are not universal.
Support to the factorisation theorem has been provided by analyses of diffractive dijet cross sections in DIS~\cite{Aktas:2007bv,Aaron:2010su,Chekanov:2007aa}. 
However, hard scattering factorisation was proven to fail at the Tevatron~\cite{Aaltonen:2012tha,Aaltonen:2010qe} and LHC~\cite{Chatrchyan:2012vc}. In resolved photoproduction at HERA, where the exchanged real photon behaves like a hadron, factorisation is expected to fail as in the hadron--hadron case. Whether H1 and ZEUS dijet photoproduction data show a suppression along the range of $x_\gamma$, the fraction of the photon momentum entering the hard scattering, as predicted by theory~\cite{Kaidalov:2001iz,Kaidalov:2003xf}, has been a dilemma for the last decade. H1 measures a suppression factor of around 0.6~\cite{Aaltonen:2012tha,Aktas:2007bv}, independent of $x_\gamma$, whereas the ZEUS data~\cite{Chekanov:2007aa} are consistent with no suppression. To shed light on this difference is highly desirable. This can be achieved only with a ZEUS analysis of the post-upgrade data. Another handle might come from HERA II data on diffractive charm photoproduction -- the HERA I data were analysed~\cite{Chekanov:2007pm,Aktas:2006up} but the large stastistical uncertainty was a limiting factor. Gap-suppression effects were seen also in the photoproduction of jet pairs separated by a LRG~\cite{Chekanov:2006pw}; consensus was not reached with theorists on their interpretation, but certainly this is a production process fertile for further investigation.

\vspace{0.3cm}
Guidelines for the {\bf measurements never carried out before} are

\noindent
- {\it high-statistics opportunities}, which mostly apply to H1 post-upgrade data and notably to the proton-tagged data (which, due to the severe acceptance limitations, are statistically disfavoured in the HERA\,I samples). Top of the list is the extraction of $F_2^c$ and $F_2^b$ in diffraction. The former was already measured with the HERA I data tagging the rapidity gap~\cite{Chekanov:2007pm,Aktas:2006up}. Both would be valuable ingredients for global fits. 

\noindent
- {\it interpreting the data where not already done before}. In addition to the 
investigation of charm, beauty or dijet processes in diffraction and the measurement 
of the related contribution to the diffractive structure function, another lens 
on the proton structure is provided by final states with a leading neutron or a 
leading proton, described in terms of the exchange of ``sub-leading'' trajectories, dominantly the pion and the Reggeon. 
The production of leading neutrons may therefore provide constraints on the structure 
of the pion at low- to medium-Bjorken-$x$ values, while fixed target experiments are limited to higher $x$ values. 
$F_2^\pi$ was extracted \cite{Chekanov20023,Aaron:2010ab}, but a QCD analysis was 
never perfomed. The high statistics H1 leading neutron HERA II data could be used as 
input to a pion parton density extraction. 

\noindent
- {\it seeking very low cross section processes}. Central exclusive production processes are the most sought exclusive processes in proton--proton and proton--antiproton collisions. Evidence was found at TEVATRON and LHC (see e.g.~\cite{Aaltonen:2015uva,Chatrchyan:2012vc} and references therein). At HERA such a 
topology is realised in double Pomeron exchange processes like $\gamma^*p \to \rho_0 Xp$, with rapidity gaps on both sides of the central system $X$. 
For tagged photoproduction the typical $M_X$ range would be 
1--6\,GeV, with the maximum at 2--3\,GeV, therefore very intriguing for the observation of low-mass resonances (glueballs?).  It is not yet established if the experimental coverage in rapidity is enough to incorporate the two gaps and to measure $M_X$ at the same time -- even a feasibility study would be interesting. 

\vspace{0.3cm}
There are {\bf measurements worth repeating} with the augmented statistics now available in the field of exclusive vector meson production. H1 and ZEUS are presently finalising their analyses of the post-upgrade data, notably for low cross section processes.  
Concerning a global interpretation of the available vector meson data, there might be potential studies left behind. For instance, H1 undertook a global fit of H1, ZEUS and OMEGA data on elastic $\rho^0$ production for the extraction of the Pomeron trajectory, but this work is still preliminary~\cite{h1prel-rhot}. It would be highly desirable to bring it to completion with the addition of the latest ZEUS data~\cite{Chekanov:2007zr} and of the low-energy data.
It must be said that the high-$|t|$ domain, very suitable to search for parton dynamics driven by BFKL evolution, has been little explored so far. HERA I data on light vector mesons, $J/\psi$ and real photons at high$-|t|$ have allowed tests of BFKL predictions but their limited 
precision has prevented a firm interpretation. Looking at the HERA\,II data 
would greatly improve the precision and hence strengthen the conclusions.

%% file: Contributions/LeszekMotyka/motyka.tex
\subsection{Twist decomposition in DIS and DDIS in the dipole approach\\
{\it L. Motyka}
}

The standard QCD description of the proton structure functions in DIS and diffractive DIS based on the leading twist-2 DGLAP evolution scheme is usually quite precise and successful. However, this approximation has limitations. In the complete description of the proton structure one should take into account also the higher twist terms in the Operator Product Expansion. From a pragmatic point of view the higher twist contributions may introduce a discrepancy between the DGLAP fits and the data and contribute to errors and uncertainties of parton densities. In particular the relative magnitude of the higher twist terms may be used to determine the applicability range of the DGLAP description. Although the higher twist components are suppressed by powers of $Q^2$ the most important higher twist corrections at HERA at small~$x$ are enhanced by additional powers of a large gluon density $xg(x,Q^2)$. So the small $x$ and moderate $Q^2$ regime are sensitive to higher twist effects.

On the other hand the higher twist components of the proton structure provide new and unique information on proton structure and QCD evolution. In particular in higher twist measurements one probes multi-parton distributions and correlations. These objects are related to multiple scattering effects and governed by higher twist evolution equations. Currently, neither the theoretical understanding nor experimental information on higher twist components of the proton structure are satisfactory and new input here should be enlightening.

Theoretical studies of higher twist effects in DIS \cite{Bartels:2000hv,Bartels:2009tu} and DDIS \cite{Motyka:2012ty} data at small~$x$ were performed within the QCD inspired colour dipole model with saturation (the GBW model) \cite{GolecBiernat:1998js,GolecBiernat:1999qd,Bartels:2002cj}. Recently, also another approach to higher twist effects at small~$x$ was proposed \cite{Motyka:2014jpa} based on the BFKL/BK small~$x$ resummation formalism, with a significantly different higher twist pattern than emerging from the saturation model. The most important finding of these studies is an evidence of strong (of the order of 50\%) positive higher twist effects in diffractive DIS at small $x_P$  and for $Q^2$ below 5~GeV$^2$. The saturation model predicts correctly the $(x,Q^2)$ DGLAP breakdown line in DDIS and provides a much better description of the DDIS data than the leading twist DGLAP fit~\cite{Motyka:2012ty} .

In inclusive DIS the sensitivity to higher twist effects is smaller than in diffractive DIS. The highest higher twist corrections are expected in $F_L$ -- about 20\% at $Q^2 = 2$~GeV$^2$ and $x=3\cdot 10^{-4}$, in $F_2$ one predicts a cancellation of higher twist effects from $F_L$ and $F_T$ and a small few per cent effect down to $Q^2 =1 $~GeV$^2$. Thus in the preceisely measured `reduced cross section', $\sigma_r = F_2 - \frac{y^2}{1 + (1-y)^2}F_L$, one should expect sizeable higher twist effects at larger~$y$. Indeed, a preliminary analysis of the combined HERA data for $\sigma_r$ shows some problems of the DGLAP fits at small~$x$ and $Q^2$ \cite{dis14talkvoica}.

Thus, the combined HERA data are expected to reveal the higher twist effects in diffractive and inclusive DIS. In order to determine them with the best precision, fits to the DIS and DDIS reduced cross sections should be performed that will combine the standard DGLAP evolution at leading twist and non-leading twist corrections. The latter are known or may be computed within saturation models or assuming a significantly different BFKL/BK scheme, and the fits may be able to differentiate between these models of high energy scattering. Another promising possibility in the twist analysis with less model dependence would be to apply a generic parameterization for the twist-4 correction, based on its $Q^2$ scaling. Such studies are expected to provide a stronger evidence for HERA findings of higher twist effects at small~$x$, more accurate parton densities and a valuable insight into the multiple scattering mechanism in QCD.

%% file: Contributions/MichelSauter/sauter.tex
\subsection{Photoproduction of $\pi^+ \pi^-$ pairs in a model with tensor-pomeron and vector-odderon exchange \\
{\it M. Sauter, A. Bolz, C. Ewerz, M. Maniatis, O. Nachtmann, A. Sch\"{o}ning}
}

We consider a Monte Carlo event generator for the reaction $\gamma p \rightarrow \pipi p$ at HERA
energies, which is based on the recently formulated model for soft high-energy scattering~\cite{Ewerz:2013kda}.
In this model, the pomeron and the $C=+1$ reggeons are formulated as effective tensor exchanges, and the 
odderon and $C=-1$ reggeons as vector exchanges. In the high energy limit the same results are obtained as for 
the standard Donnachie-Landshoff pomeron. The model provides everything to apply it to a concrete
calculation of amplitudes. Namely, it provides a large set of Feynman rules for vertices and propagators (pomeron, odderon and reggeons) 
and a list of form factors. It also includes photon-exchanges (Primakoff-effect) formulated in the vector-dominance model, and 
estimated values of all known parameters. 

\begin{figure}[thb]
\center \includegraphics[scale=0.72]{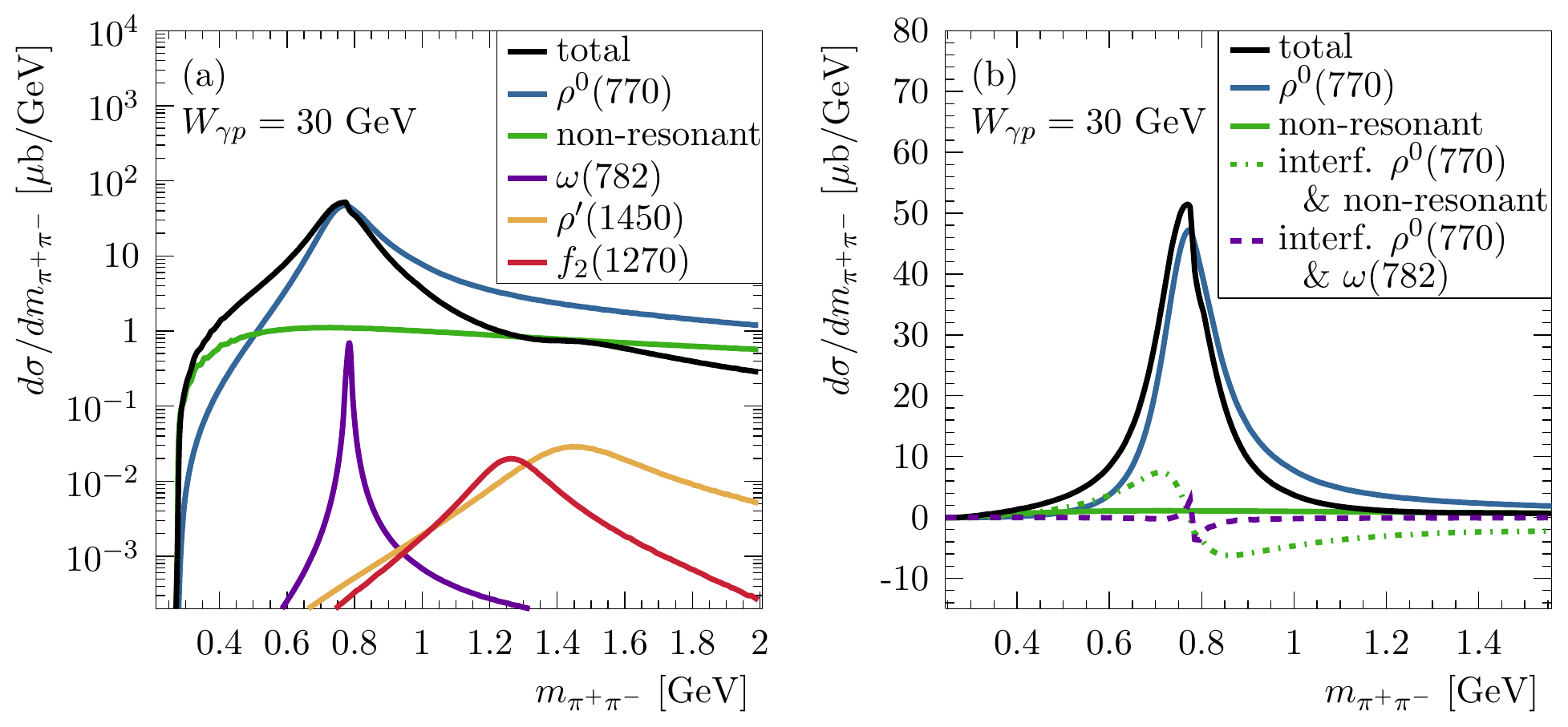}  
\caption{Differential cross sections 
$ d\sigma / d \mpipi \left( \gamma p \rightarrow \pipi p \right)$ as
  function of $\mpipi$ for fixed $\Wgp = 30 \gev$ and integrated over the
  range $-1 \gevsq \leq t \leq 0$.
(a) The full model, non-resonant contributions and the contributions from the resonances 
$\rho^0(770)$, $\omega(782)$, $f_2(1270)$ and $\rho'(1450)$ are shown.
(b) Dominant  contributions in the
  $\rho$ mass region 
including the leading interferences of $\rho^0(770)$ with the non-resonant
$\pipi$ production and the
$\omega(782)$ meson are shown. 
\label{fig: 1}}
\end{figure}

\begin{figure}[thb]
\center
\setlength{\unitlength}{1.0cm}
 \includegraphics[scale=0.72]{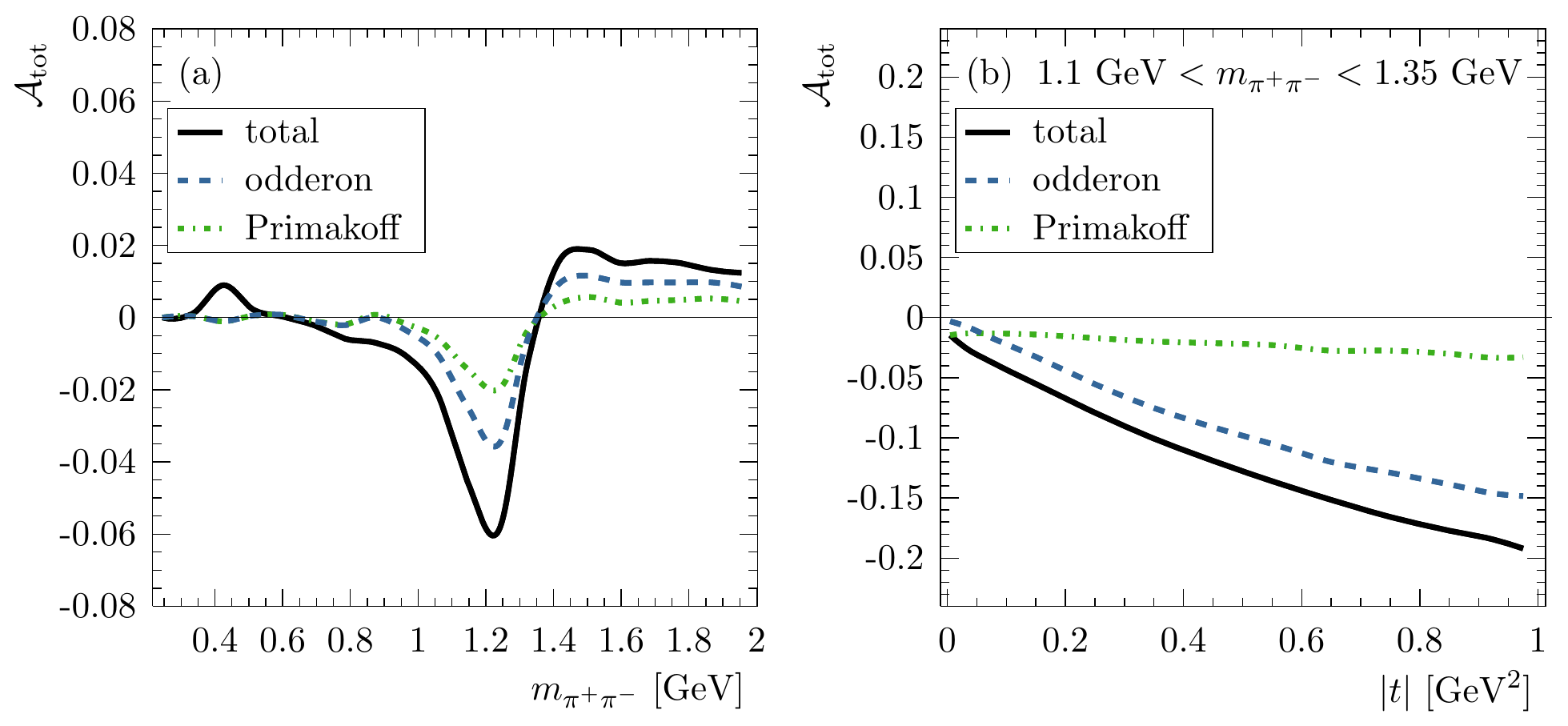}  
\caption{Total charge asymmetry ${\cal A}_\textrm{tot}$ as function of the
  (a) invariant mass of the $\pipi$ system and (b) squared momentum transfer $t$.
The asymmetries are presented for fixed $\Wgp = 30 \gev$ and (a) integrated over
 the range 
$-1 \gevsq \leq t \leq 0$ and (b) integrated over the range $1.1 \gev < m_{\pipi
}<1.35 \gev$.
The individual contributions to the asymmetries 
from photon (Primakoff) and odderon exchange are shown by the green
  dashed-dotted and blue dashed 
lines, respectively.
\label{fig: 2} }
\end{figure}

In the Monte Carlo event generator  published in~\cite{Bolz:2014mya}, we study dipion production via  resonances
$\rho$, $\omega$, $\rho'$, $f_2$,  via non-resonant mechanisms through the exchanges of
photon, pomeron, odderon and reggeons. The cross section as a function of dipion invariant mass $\mpipi$ is shown in
Fig.~\ref{fig: 1}(a) for the full model and the various contributions. Cross sections as a function 
$t$, $\Wgp$ and $\cos \theta_{\pi^+}$ have been shown
in the presentation~\cite{sauter} and compared to data where possible.  Further, several interference-effects
occurring between the different contributions to the total  amplitude have been discussed. 
The largest effect arises due to interference of resonant $\rho$ production with non-resonant diagrams
(Drell-S\"oding mechanism) and leads to a distortion of the $\rho$~line-shape (often referred to as "skewing-effect"),
as shown in Fig.~\ref{fig: 1}(b). Also shown is the interference of the $\rho$ with the $\omega$ meson.
Another discussed interference effect is due to a completely different nature: On the $\fTwo$ mass peak
at $1.27\gev$ the interference of $C=+1$ and $C=-1$ diagrams leads to an asymmetry 
in the angular distribution of the pions. This asymmetry can be exploited to construct a total charge asymmetry ${\cal A}_\textrm{tot}$,
shown in Fig.~\ref{fig: 2}(a) as a function of $\mpipi$. 
It has a prominent structure around $\mpipi =1.27\gev$ and is predicted to be of the order of $5-10\%$.
The total charge asymmetry is mainly driven by the interference of the high mass tail of the
$\rho$-resonance with the $\fTwo$-resonance, whereas the production of the $\fTwo$-meson is modelled by
exchanges of photons and (possible) odderons are taken into account.
The overall share of the total asymmetry is of the same order for the two contributions, but they have a different behaviour
as a function of squared momentum transfer $t$, as presented in Fig.~\ref{fig: 2}(b). 
The asymmetry due to the Primakoff-effect (photon exchange) shows as a function 
of $t$ an almost flat behaviour. In contrast the asymmetry due to a possible odderon 
increases, in absolute value, with $\vert t \vert$, reaching around
$20\%$ at $\vert t \vert = 1\gev$ for our choice of parameters. The observation 
of such a structure in data would be a clear sign for an odderon at work.

Interested readers should also consult the following related works~\cite{Ginzburg:2002zd, Lebiedowicz:2014bea, Lebiedowicz:2014wka}
and the references therein.

%% file: spin.tex
\section{Spin physics}
\label{sec:spin}

\input{Contributions/PeterKroll/kroll.tex}

\input{Contributions/AlessandroBacchetta/bacchetta}

\input{Contributions/EmanueleNocera/nocera}

%% file: Contributions/PeterKroll/kroll.tex
\subsection{GPDs from exclusive meson leptoproduction\\
{\it P. Kroll}
}

In the generalised Bjorken regime of large $Q^2$ and large $W$ but fixed
Bjorken-$x$ the amplitudes for exclusive leptoproduction of mesons and 
photons factorise in hard partonic subprocesses and generalised parton 
distributions (GPDs). For an extraction of the 
GPDs from experimental data in analogy to that of the usual parton 
distributions,  
a parametrisation of the GPDs is required. A frequently employed concept 
is to construct the GPDs from
a double distribution where the latter is represented by a product of a
zero-skewness GPD and a weight function that generates the skewness dependence.
The zero-skewness GPDs is parametrised as their $\xi=t=0$ limit (i.e.\
a PDF for $H$ or $\widetilde H$, for others parametrised like a PDF) multiplied by 
an exponential in $t$ with a Regge-like profile function, for a review see 
\cite{kroll} and references therein.
 
From the form factors of the proton which represent the lowest moment of
appropriate GPDs, the GPDs $H, E$ and $\widetilde H$ for valence quarks are
determined and used in subsequent analyses. The longitudinal
cross section for $\rho^0$ and $\phi$ leptoproduction fix $H$ for sea quarks
and gluons, HERMES data on $\pi^+$ electroproduction provide information on 
$\widetilde H$ and the transversity GPD $H_T$. Spin
density matrix elements for $\rho^0$ and $\omega$ production as well as
JLAB data on $\pi^0$ production hint at a large contribution from $\bar{E}_T$
which has also been observed in a lattice QCD calculation. This 
result however
needs confirmation. Data on the $\pi^0$ cross section from HERMES or COMPASS
would allow for a determination of $\bar{E}_T$. Data on $A_{LU}(\pi^0)$ would
also be of help.  

Not much is known on the GPD $E$ for gluon and sea quarks as yet. Their 
contributions to the $\sin{(\phi -\phi_s)}$ modulation of the transverse target
spin asymmetry measured by HERMES and COMPASS, being related to an
interference term of $E$ and $H$, cancels to a high degree
of accuracy. However, the corresponding observable for DVCS, although 
suffering from rather large errors, tells us at least that the convolution of
$E_{\rm sea}$ is likely negative. This rather poor information
in combination with a positivity bound for $E_{\rm sea}$ allows for an 
evaluation of the angular momenta the partons inside the proton carry.
At a scale of 2\,GeV the spin of the proton is shared about equally  by
$u$ quarks and gluons. In the context of the much debated  so-called 
spin crisis this is an interesting result. A measurement of $A_{UT}$ for
DVCS with substantially smaller errors would be extremely helpful in reducing 
the uncertainties of the angular momenta.

%% file: Contributions/AlessandroBacchetta/bacchetta.tex
\subsection{Inclusive and semi-inclusive spin physics\\
{\it A. Bacchetta}
}

The experimental 
investigation of the internal structure of the nucleon has been the
``core business'' of the HERA experiments. The ultimate goal of this investigation
is to produce maps of the distribution of partons in the nucleon in all
possible details. Why do we map partons? The main reasons are: to understand
QCD and to make predictions for hadronic collisions. In the
last years, much emphasis was put on the
second motivation, but the first should not be overlooked. From this point
of view, we should not focus only on specific corners of these maps, but widen
our view as much as possible. 

Unpolarised standard PDFs represent a very important specific corner of
nucleon mapping. HERA experiments have given and keep giving landmark
contributions to the reconstruction of PDFs.  
They can be intuitively viewed as partonic maps in one
dimension. A way to widen the view is to map partons in
more dimensions, 
in particular accessing the so-called Transverse Momentum
Distributions (TMDs) in semi-inclusive DIS. 
The HERMES experiment at HERA truly pioneered this exploration, 
even though it was not foreseen in the original plans. The HERMES
Technical Design Report~\cite{HERMES:1993aa} mentioned several goals, all related to
collinear (polarized) PDFs, with no mentioning of TMDs. However, the most
cited paper of the collaboration to this date is related to
TMDs~\cite{Airapetian:2004tw}.   

Given the relative novelty of 3D mapping, it is still possible to find
measurements to be performed using the unique data from HERA. 

Let us start first from unpolarized measurements, where all three HERA 
experiments can contribute. The formula for the semi-inclusive DIS cross
section, differential in all possible variables, is (see
\cite{Bacchetta:2006tn} for the
definitions of all elements involved) 
\begin{equation} 
\begin{split} 
\frac{d\sigma}{d\xbj \, dy\, d\psi \,dz\, d\phi_h d P_{h\perp}^2}
& = 
\frac{\alpha^2}{\xbj y  Q^2}\,
\frac{y^2}{2(1-\varepsilon)}\,  \biggl( 1+\frac{\gamma^2}{2\xbj} \biggr)\,
\bigg\{
F_{UU ,T}
+ 
\varepsilon
F_{UU ,L}
\\ &
+ \sqrt{2\varepsilon (1+\varepsilon)}\cos\phi_h\,
F_{UU}^{\cos\phi_h}
+ \varepsilon \cos(2\phi_h) 
F_{UU}^{\cos 2\phi_h} \bigg\}.
\label{e:unpolSIDIS}
\end{split} 
\end{equation}  
When integrated over the outgoing particle, the formula reduces to the
well-known inclusive DIS formula with structure functions $F_T$ and $F_L$.

HERMES analysed all terms of the above formula. The multiplicities,
proportional to the structure function $F_{UU,T}$, have been published in
Ref.~\cite{Airapetian:2012ki}. A lot of information about TMDs 
can be obtained from these measurements
(see, e.g.\ \cite{Signori:2013mda,Anselmino:2013lza}), 
although other measurements are required. 
HERMES could still analyse $\pi^0$
and $\eta$ production, and some data from 2006/2007 still need to be included
in the analysis. ZEUS and H1 measured the standard structure
functions  $F_T$
and $F_L$, but also 
some measurements with transverse-momentum
dependence~\cite{Alexa:2013vkv}.  
However, it would be very interesting to make an analysis
similar to HERMES, especially because the extension of the $Q^2$ and $x$ range
is crucial to gain insight into the properties of TMDs~\cite{Collins:2011zzd}, 
in particular their
evolution
equations~\cite{Aybat:2011zv,Anselmino:2012aa,Aidala:2014hva,
Echevarria:2014xaa}, 
currently under careful scrutiny.

The cosine modulations in Eq.~\eqref{e:unpolSIDIS} contain also twist-3
contributions. Measurements of these modulations have been performed by
HERMES~\cite{Airapetian:2012yg} and 
ZEUS~\cite{Breitweg:2000qh,Chekanov:2006gt}.
They are more
complicated and have a less transparent interpretation compared to twist-2
contributions, but they are also a source of new information about QCD, not
accessible at twist-2, and are largely unexplored. 
Moreover, structure functions that
are dominated by twist-2 contributions always contain some higher-twist
contaminations. Therefore, high-precision extractions of twist-2 observables
requires 
the knowledge of the twist-3 contribution (see, e.g.\ the recent discussion in
Ref.~\cite{Jimenez-Delgado:2013boa}).
 
On top of this, it would be interesting also to look for signals that are not
included in the above formula. They can be due to parity-violating electroweak
interactions, or to two-photon exchange~\cite{Airapetian:2009ab}. 
It would be much less likely 
and much more surprising to be able to spot physics beyond the Standard
Model (e.g.\ strong parity violation or heavy photons).

Let us turn now to polarised measurements. First of all, we could include
lepton polarisation and obtain an additional $\sin \phi_h$ modulation in the
cross section of Eq.~\eqref{e:unpolSIDIS}. 
This modulation arises from twist-3 contributions.  
If we include also target polarisation (something that only
HERMES can do), the situation becomes more complex, with 18 different structure
functions~\cite{Bacchetta:2006tn}. HERMES has published measurements of some
of them, without a full multi-dimensional
binning~\cite{Airapetian:2004tw,Airapetian:2005jc,Airapetian:2006rx,
Airapetian:2009ae,Airapetian:2010ds,Airapetian:2011wu}.  
The ``ultimate'' HERMES polarised DIS
analysis (multi-dimensional, with all structure functions) is in progress, but
still needs to be published. 

Apart from single-hadron DIS measurements, also two-hadron
measurements~\cite{Gliske:2014wba}  and
single-hadron inclusive leptoproduction should be taken into
consideration. HERMES has already published something on these
topics~\cite{Airapetian:2008sk,Airapetian:2013bim}, but
something still remains to be done. ZEUS and H1 could try similar
analyses in the unpolarised sector. 

Apart from DESY experiment, similar measurements have been performed by 
the COMPASS collaboration~\cite{Alexakhin:2005iw,Ageev:2006da,Alekseev:2008aa,Alekseev:2010rw,Alekseev:2010dm,Adolph:2012nw,Adolph:2012sp,Adolph:2012sn,Adolph:2012nm} and Jefferson Lab
collaborations~\cite{Qian:2011py,Huang:2011bc,Avakian:2010ae,Aghasyan:2011ha,
Allada:2013nsw,Zhang:2013dow}. 
More measurements will come in the near future from these
experiments. In the far future, an Electron Ion Collider 
would be an excellent
machine to fully exploit the rich opportunities of this
field~\cite{Boer:2011fh,Accardi:2012qut}.

%% file: Contributions/EmanueleNocera/nocera.tex
\subsection{The longitudinal spin structure of the nucleon\\
{\it E. Nocera}
}
\label{sec:collinearspin}

The decomposition of the one-half nucleon spin into individual contributions 
arising from the spin and orbital angular momentum of quarks and gluons is 
encoded in the helicity sum rule~\cite{Leader:2013jra}, which may be written 
as~\cite{Jaffe:1989jz}
\begin{equation}
\frac{1}{2} 
= 
\frac{1}{2}\Delta\Sigma(\mu^2) + \Delta G(\mu^2) 
+ 
\mathcal{L}_q(\mu^2) + \mathcal{L}_g(\mu^2)
\,\mbox{,} 
\label{eq:helicitysumrule}
\end{equation}
with $\Delta\Sigma$ and $\Delta G$ the first moments of the helicity-dependent,
or polarised, parton distribution functions
\begin{equation}
\Delta\Sigma(\mu^2) 
= 
\sum_{q=u,d,s}\int_0^1 dx\,\left[\Delta q(x,\mu^2) + \Delta \bar{q}(x,\mu^2)\right]
\ \ \ \ \ \ \ \ \ 
\Delta G (\mu^2) 
=
\int_0^1 dx \,\Delta g(x,\mu^2)
\,\mbox{,}
\label{eq:moments}
\end{equation}
and $\mathcal{L}_{q,g}$ the quark and gluon orbital angular momentum. All these
quantities depend on the factorisation scale $\mu$, and such a dependence can 
be computed in perturbative Quantum Chromodynamics (QCD). Polarised parton 
distributions are defined as the net amount of momentum densities of partons 
polarised along or opposite the polarisation direction of the parent nucleon.
Because of Eq.~\eqref{eq:moments}, they encode the information on the 
longitudinal spin structure of the nucleon, and much effort has been devoted 
to determine their non-perturbative dependence on the momentum fraction $x$ carried
by partons in several global analyses of experimental data (for a review, see
e.g.\ Ref.~\cite{Nocera:2014vla}).

The bulk of the experimental information on polarised distributions comes from
neutral-current inclusive and semi-inclusive deep-inelastic scattering (DIS 
and SIDIS) with charged lepton beams and nuclear targets. The HERMES experiment
at DESY has significantly contributed to pioneering measurements of this kind,
specifically on proton, deuteron and neutron DIS structure functions 
$g_1^{p,d,n}$~\cite{Ackerstaff:1997ws,Airapetian:1998wi,Airapetian:2006vy}
and $g_2^{p}$~\cite{Airapetian:2011wu}, and on SIDIS spin-asymmetries 
for the production of $\pi^\pm$, $K^\pm$ and charged 
hadrons~\cite{Ackerstaff:1999ey,Airapetian:2003ct,Airapetian:2004zf}.
Because of the way the corresponding observables factorise, inclusive DIS data 
constrain the total quark combinations $\Delta q^+=\Delta q+\Delta\bar{q}$, 
$q=u,d,s$, while hadron-tagged SIDIS data constrain individual quark and 
antiquark distributions. In principle, both DIS and SIDIS data would also
constrain the gluon distribution $\Delta g$ via scaling violations, but in 
practice their effect is rather weak because of the small $Q^2$ range covered. 
One- or two-hadron production may provide a direct handle on $\Delta g$, and 
this has also been measured by 
HERMES~\cite{Airapetian:1999ib,Airapetian:2010ac}. 

The inclusion of SIDIS data in a global analysis of polarised parton 
distributions requires an accurate knowledge of fragmentation functions: 
indeed these enter the factorised expression of the corresponding observables,
since they encode the non-perturbative fragmentation of partons into the 
observed hadrons.
Fragmentation functions are on the same footing as parton distributions, hence
they must be determined from a global analysis of experimental data, see 
e.g.\ Ref.~\cite{Albino:2008gy}. The HERMES experiment also provided data
useful for such a determination, specifically multiplicities for $\pi^0$, 
$\pi^\pm$ and $K^\pm$ production in 
SIDIS~\cite{Airapetian:2001qk,Airapetian:2012ki}.
These were recently included in a global QCD analysis of $\pi^\pm$ fragmentation
functions~\cite{deFlorian:2014xna}, and it was found that they are in fairly 
good 
agreement with a large variety of different measurements in $e^+e^-$ and $pp$
collisions. These HERMES data are presented in both $[x,z]$ and $[Q^2,z]$ bins,
where $x$ is the target momentum fraction carried by the parton, $z$ is the 
final hadron momentum fraction carried by the parton, and $Q^2$ is the 
energy transferred in the process. It was shown~\cite{Leader:2013kra} 
that $[x,z]$ and $[Q^2,z]$ presentations of HERMES data are not mutually 
consistent, and that $\pi^\pm$ $[Q^2,z]$ data (also used in 
Ref.~\cite{deFlorian:2014xna}) may be not consistent with analogous preliminary
data measured by the COMPASS experiment~\cite{Makke:2013bya}.
Also, it has been advocated that the HERMES data bin corresponding to the 
smallest value of $z$, $0.2\leq z \leq 0.3$ lies in a kinematic region where 
the usual QCD description of SIDIS may be not reliable~\cite{Leader:2014oxa}.
Further careful phenomenological studies, possibly including a re-analysis of
HERMES multiplicities and their inclusion in a methodologically improved global
determination of fragmentation functions, are needed to clarify this state 
of affairs.

Overall, the potential of the HERMES experiment in unveiling the longitudinal
structure of the proton has been fully exploited, with all possible analyses
finalized. Unfortunately, these allow for a determination of $\Delta\Sigma$ 
and $\Delta G$, Eq.~\eqref{eq:moments}, which remains largely uncertain and 
prevents from any firm conclusion on their contribution to 
Eq.~\eqref{eq:helicitysumrule}~\cite{Ball:2013lla}. 
This is especially due to the dominance of 
the uncertainty coming from extrapolation in the small-$x$ region,
$x\lesssim 10^{-3}$, not covered by experimental data. In order to 
overcome this issue, a significantly extended kinematic coverage and increased
statistical precision, in comparison to the HERMES reach, are needed. 

For instance, a significant amount of data in longitudinally polarised 
proton--proton ($pp$) collisions have become available from the 
Relativistic Heavy Ion Collider (RHIC)~\cite{Aschenauer:2015eha}, 
specifically single-spin asymmetries for $W^\pm$ 
production~\cite{Adamczyk:2014xyw} and longitudinal double-spin asymmetries 
for $\pi^0$ and high-$p_T$ inclusive jet 
production~\cite{Adare:2014hsq,Adamczyk:2014ozi}. These data have been 
included in two independent global analyses of longitudinally polarised parton
distributions, {\tt DSSV14}~\cite{deFlorian:2014yva} and 
{\tt NNPDFpol1.1}~\cite{Nocera:2014gqa}, and  
provided first evidence respectively of a sizeable, positive polarised light 
sea quark asymmetry, $\Delta\bar{u}-\Delta\bar{d} > 0$, and of a sizeable, 
positive polarised gluon distribution, $\Delta g > 0$, though in the 
limited range of momentum fractions covered by RHIC,
$0.04\lesssim x \lesssim 0.4$. Further improvements of these results are 
expected after RHIC run-2015~\cite{Aschenauer:2015eha}.

In order to probe the nucleon spin structure at the intensity and energy 
frontier, brand new facilities will be required. On the one hand, the $12$ GeV
Jefferson Lab (JLab) electron beam energy upgrade~\cite{Dudek:2012vr}
will extend the kinematic reach of the existing JLab data
to twice smaller $x$ as well as to larger $x$ values: large luminosity and high 
resolution available will allow for a substantial reduction of parton 
distribution uncertainties in the medium-to-large $x$ region, and for a 
discrimination among different models of nucleon 
structure~\cite{Nocera:2014uea}. On the other hand,
a high-energy, polarised Electron--Ion Collider (EIC)~\cite{Accardi:2012qut}
will extend the kinematic reach down to $x\sim 10^{-4}$ and up to 
$Q^2 = 10^4$ GeV$^2$:
this  will allow for an accurate determination of $\Delta g$ via scaling 
violations in inclusive DIS, of $\Delta\bar{u}$ and $\Delta\bar{d}$
via inclusive DIS at high $Q^2$ mediated by electroweak bosons,
and of $\Delta s$ via kaon-tagged 
SIDIS~\cite{Aschenauer:2012ve,Ball:2013tyh,Aschenauer:2013iia}.
Also, would the EIC data confirm the {\tt DSSV14} or {\tt NNPDFpol1.1} best fit
behaviours of $\Delta\Sigma$ and $\Delta G$, only a small fraction of the proton 
spin is expected to come from orbital angular momentum, 
$\mathcal{L}_q + \mathcal{L}_g$ in 
Eq.~\eqref{eq:helicitysumrule}~\cite{Aschenauer:2015ata}.

%% file: montecarlo.tex
\section{Monte Carlo programmes for HERA physics}
\label{sec:montecarlo}

\input{Contributions/FrancescoHautmann/hautmann.tex}

\input{Contributions/SimonPlaetzer/plaetzer.tex}

%% file: Contributions/FrancescoHautmann/hautmann.tex
\subsection{Parton shower  Monte Carlo  generators beyond collinear approximations\\
{\it F.~Hautmann, H.~Jung}
}

Phenomenological  studies of  final states  at hadron  
colliders use 
Monte Carlo event generators~\cite{Buckley:2011ms}  
based on 
collinear evolution of QCD parton showers combined 
with perturbative hard matrix elements. 
It has been known for a long time, however, that 
when    this  
 approach   is pushed  to increasingly high 
 energies     new effects 
 arise from non-collinear corrections to parton 
 branching processes~\cite{Ciafaloni:1987ur},  
  due to soft  but       finite-angle  
   multi-gluon  emission. 
As  was noted already long 
ago~\cite{Marchesini:1992jw},  
 such   effects  
 can be taken into account   by  
treating the QCD evolution of the  initial 
 state   via       transverse-momentum 
dependent  branching algorithms  
  coupled~\cite{Catani:1990eg}  to hard matrix 
elements at  fixed transverse momentum.  

Besides these dynamical  contributions, 
effects of kinematical origin have more recently 
 been  pointed  out~\cite{Hautmann:2012dw}  
 which arise from  combining  the collinearity 
approximations in the  showering 
algorithms  with energy--momentum 
conservation constraints.  These effects  are 
  responsible   for a  significant  
fraction of the large  
parton-showering corrections~\cite{Dooling:2012uw} 
observed for   a number  of 
 hadronic final-state observables at the LHC, 
particularly  at  noncentral rapidities 
of jets~\cite{Deak:2011ga},  photons  and  leptons~\cite{Hautmann:2012sh}.

Parton-shower algorithms which go beyond 
the collinear approximations  are thus needed 
for precision phenomenology at high-energy 
colliders, 
if one is to evaluate the theory consistently and 
control the theoretical uncertainties associated 
with higher-order QCD radiation. 
DIS measurements play an important role in this. 
An example  is given  in~\cite{Dooling:2014kia}. 

Ref.~\cite{Dooling:2014kia}
 studies the effects of soft-gluon 
coherence from   finite-angle multiple QCD  
emission~\cite{Ciafaloni:1987ur,
Marchesini:1992jw,Hautmann:2008vd,Hautmann:2014uua} 
 on   multi-jet final states associated with Drell-Yan 
vector boson production at the LHC. 
These effects    go beyond any calculation 
based on  next-to-leading-order
perturbation theory matched with 
collinear parton showers (see 
e.g.~\cite{Hoeche:2012ft})  but according to 
the  method~\cite{Dooling:2014kia}  
can be determined via 
 high-energy factorisation~\cite{Catani:1990eg}  
from  deep inelastic scattering. 
The results of~\cite{Dooling:2014kia} incorporate 
the high-precision DIS data as described 
in~\cite{Jung:2014vaa},   and  show  that 
multi-gluon emissions {\em beyond  next-to-leading 
order} and {\em beyond collinear logarithmic 
 approximations}    
 are extremely relevant to ascribe theoretical 
uncertainties to $W$ + jets predictions. 

This in turn implies that new determinations  of 
parton distribution  functions to be used in 
conjunction with 
parton showers are  needed, in order to 
consistently take into account  important 
radiative effects in higher order, 
 and that the 
use of  combined  DIS and Drell-Yan  data  has  
the  potential to  constrain  such determinations.  

 DIS   structure function 
measurements as well as 
 single-particle and jet cross sections will be 
relevant for this.     A particularly 
 interesting -- and very 
challenging  -- 
area  will involve  charged-particle distributions  
 in the region of low $p_T$.  In this region, 
 connecting single-particle cross sections  
at HERA and 
 mini-jet (or leading-particle) production 
at the LHC~\cite{Grebenyuk:2012qp} 
 could contribute  to theoretical  insights  into 
the   physics of  parton saturation.

%% file: Contributions/SimonPlaetzer/plaetzer.tex
\subsection{Herwig++ for $ep$\footnote{Work presented on behalf of the Herwig++ collaboration.} and 
Rivet for $ep$\footnote{Work in progress with Hannes Jung.}\\
{\it S. Pl\"{a}tzer}
}

\noindent {\bf Herwig++} \cite{Bahr:2008pv} builds on the experience
gained with its predecessor HERWIG, but features a number of new
capabilities and improvements to all aspects of the
simulation. Being designed mainly with the LHC in mind, we
have, however, always emphasised the importance of legacy and future
collider data. Not only is this required for a thorough validation and
tuning of the non-perturbative models considered in Herwig++
\cite{Bahr:2008dy,Gieseke:2012ft}, but also for improvements to the
description of the hard scattering process and subsequent parton
showering. In its current release 2.7 \cite{Bellm:2013lba}, Herwig++
offers two parton shower modules: the default angular ordered parton
shower based on \cite{Gieseke:2003rz}, and a dipole-type parton shower
described in more detail in \cite{Platzer:2009jq,Platzer:2011bc}. Both
parton showers are able to simulate deep inelastic scattering without
any restriction; the multiple interactions module is not yet ready to
address photoproduction though there is no conceptual obstacle in
doing so. 

Next-to-leading order QCD corrections to the hard scattering process
have been considered for a variety of processes and have been matched
to both parton cascades available. This has mainly been done for the
angular ordered shower on a process-by-process basis following the
Powheg method, see \cite{Hamilton:2008pd,Hamilton:2009za} for
examples, and both via the MC@NLO and Powheg methods for the dipole
shower, using a more flexible framework for NLO calculations and
matching, see \cite{Platzer:2011bc} for more details. Specifically DIS
has been addressed at NLO in combination with both shower algorithms
\cite{D'Errico:2011um,Platzer:2011bc}. Based on the development of a
flexible NLO framework, the upcoming release of Herwig++ will feature
a fully-automated way of calculating NLO QCD corrections to Standard
Model processes at $pp$, $ep$ and $ee$ colliders, via both major
matching paradigms and for both shower modules available.

Besides reaching NLO accuracy for the hard process with the fewest
number of external legs, the combination (or merging) of hard
processes with different jet multiplicities and parton showers, both
at leading and next-to-leading order is desirable to obtain a smooth
description across jet bins. A modified leading order merging
algorithm has been pioneered for the angular ordered shower
\cite{Hamilton:2009ne}, and work is underway to follow up recent
proposals on next-to-leading order merging
\cite{Platzer:2012bs,Lonnblad:2012ix} with the dipole shower.

Both the NLO matched simulations and those based on merging different
jet multiplicities have so far only been studied in great detail for
LEP, Tevatron and LHC data, especially for the case where additional,
hard jets become relevant. Comparisons to HERA and future $ep$ data
are extremely important to validate these new algorithms, but have so
far not been addressed due to the lack of analyses which can easily be
integrated with LHC-age Monte Carlo event generators, notably via the
{\bf Rivet} framework \cite{Buckley:2010arxb}.

To overcome this limitation, first steps have been undertaken to
develop a Rivet wrapper around HZTOOL to provide easy access to the
analyses contained there. This Rivet plug-in is currently subject to
validation and will be made available along with an upcoming HZTOOL
release. It should be stressed, however, that this does not
encourage the addition of new analyses of HERA data into HZTOOL, but rather to
provide them directly as Rivet plug-ins, with a minimum amount of
corrections, {\it i.e.} solely account for detector acceptance,
applied to the data. 

In terms of {\bf future analyses}, certainly the characterisation of
multijet final states is highly welcome from the perspective of
contemporary event generator development, though also different jet
algorithms and jet shapes are valuable input to benchmark new shower
algorithms and improvements. If possible, data sensitive to
hadronisation effects, especially in the forward region, could help in
constraining colour reconnection models which are highly relevant to
LHC physics.

%% file: summary.tex

\input{Contributions/JohnDainton/dainton}

%% file: Contributions/JohnDainton/dainton.tex
\section{Summary: from Dirac's electron to Dirac electrons and quarks\footnote
{The presentation on which this very short summary is based is very much a personal perspective of 22 years of 
HERA physics. It is based on an invitation to speak at the end of the colloquium and workshop at DESY in November 2014.  
It is not inclusive of the multitude of results and measurements from the four HERA experiments, H1 ZEUS, HERMES and 
HERA-B. It is made possible by generations of colleagues, both on the experiments and on the HERA machine, whose 
hard work, dedication, innovative determination and unswerving commitment has secured HERA as pivotal in the 
development of 20th and early 21st century physics, culminating in the SM of today. Specific contributions to this colloquium, 
which are to be found in other presentations included with this short summary, contain more details of latest results in respect 
of what is written down here.  A long write-up is nearly complete which will be published as a DESY preprint shortly.
} 
\\
{\it J. Dainton}
\label{sec:summary}
}
{\it                                                                              
 "(The) history of science has shown that even during that phase of her progress in which she devotes herself to improving the 
 accuracy of the numerical measurements of quantities long familiar, she is preparing the materials for the subjection of new regions, 
 which would have remained unknown if she had been contented with the rough methods of her earlier pioneers."
}
 \begin{flushright}
James Clerk Maxwell
 \end{flushright}
 
 \subsection{The 1992 HERA perspective}
 
In 1992 lepton--hadron physics is recognised to be at the root of a Standard Model (SM) in which was still missing the sixth quark. 
The physics of electron--positron annihilation at LEP continues to probe with precision following discoveries at DESY and SLAC 
at lower energy which in many respects defined the template of the symbiosis of theory and experiment which gave us the SM. At 
and beyond the Fermi scale is probed with anti-proton--proton interactions at the Fermilab Tevatron with a focus on more quarks 
and as always even then on the Higgs-boson. The discovery of the SM vector-bosons is only a decade old. Hadron (proton) structure 
is maturing rapidly after the pioneering deep-inelastic experiments at SLAC, Fermilab and CERN in terms of parton density functions 
in a paradigm in which the visible valence quarks carry barely 50\% of hadron momentum. Soft hadronic physics remains a 
conundrum built round what many take to be an anachronistic and limited phenomenology based on constraints which must be met by 
scattering amplitudes -- namely a Regge asymptotic expansion. 

\subsection{HERA at the close of 2014}

HERA has discovered that hadron mass, us -- humanity and the presently visible Universe, is chromodynamic field energy. 
Less than 1\% of visible mass in the proton is attributable to constituent mass, that is to mass due to the spontaneous breaking 
of SU(2)\,$\times$\,U(1)$_{\rm broken}$, the Higgs mechanism. We are gluons! HERA has measured with precision this discovery 
such that we now have a quantitative prescription for how a proton interacts in the SM in the form of chromodynamic phenomenology. 
HERA achieved this as an ever-improving instrument which probed the structure of the proton with the unified electroweak sector, 
that is, with each, and with all together, of lepton and quark flavour, of chirality, and of the differences between lepton--matter and 
anti-lepton--matter interactions. The anti-matter--matter asymmetry of the proton is thereby measured directly for the first time.

HERA's success in quantifying hadron structure with precision reveals new issues in short-distant, dense chromodynamics, 
"low-$x$ physics", which must be met and understood in pursuing the detailed behaviour of non-abelian field theory. Remember 
that chromodynamics is the only experimentally accessible non-abelian field! The precision of the measurements, together with the 
technology used to quantify proton structure which reveals these issues, now challenges field theory with precision akin to that of 
Lamb and Rutherford who first discovered in 1947 the "Lamb-shift", who thereby opened a new window on the interactions of Dirac 
protons, and thereby whose work led the way to the SM. Taking this consequence of HERA physics onwards must be part of 
understanding unification which is inclusive of chromodynamics.

A feature of HERA physics which was completely overlooked in the years of its proposal, funding and construction, and which 
therefore now stands as another discovery, is the elucidation of the chromodynamic nature of high energy, "diffractive", hadronic, 
interaction. The ability to measure at sub-femtoscopic  inelastic, high energy, hadron production in which one proton or low mass 
baryon excitation scatters has enabled beautiful measurements of the partonic behaviour of what has been for years no more than 
a phenomenological box quantified in terms of long-range hadron--hadron interactions. The resulting "structure" of the chromodynamic 
field as it works in the form of the nucleon--nucleon interaction is now understood quantitatively with an accuracy better than that of 
the proton itself at the time of HERA's inauguration. The symbiosis of this structure of the diffractive high energy proton interaction with 
that of the structure of the proton can now, because of HERA, be clearly seen as the basis of low-$x$ physics as a cornerstone of the 
physics of dense, non-abelian, matter. Other cornerstones include heavy-ion physics at the LHC, at RHIC (BNL), and at CERN in the 
fixed target ion physics which, all together, will lead ultimately to an understanding of the phase-equilibria of confining, asymptotically 
free, chromodynamics. 

The space-like domain of electroweak, Glashow--Salam--Weinberg (GSW), unification, that is the non-abelian, 
SU(2)\,$\times$\,U(1)$_{\rm broken}$, field dynamics, is now quantified because of HERA. Unification of force at the Fermi-scale has 
been for the first time demonstrated directly to be as predicted in GSW theory. The combination of cross-section measurements of 
neutral current and charged current inclusive scattering with incident lepton and anti-lepton polarisation has revealed no evidence for 
anything unexpected in electroweak force. After 80 years Fermi's original hypothesis for a weak $\beta$-interaction is confirmed with the 
simultaneous demonstration of unification at an energy scale of $\sim$\,100\,GeV in terms of an SU(2)-isospin in respect of lepton and 
quark couplings at what we now call the Fermi energy scale. It is gratifying then that in 2012 science has finally managed to produce 
evidence with the LHC for the scalar field, the Higgs-field, which underpins the dynamic mechanism behind the SM of GSW unification, 
the Higgs-miracle, by observing its "quantum", the Higgs-boson! 

Lastly, the discoveries at HERA which must be mentioned concern both the distinction between the lepton and quark sectors of the 
SM and the properties of quarks as Dirac fermions.

Despite sensitivity to new interactions as a means of securing new physics beyond the time-like kinematic limit $\sim$\,200\,GeV, 
no evidence for anything has emerged for new physics beyond the SM at HERA. Indeed, quite the opposite. Neither direct resonant 
($s$-channel with associated phase variation with respect to SM) nor contact (or even $u$-channel space-like going beyond the 
time-like kinematic limit) (anti-)lepton--(anti-)quark phenomena have been observed. This window on one possible way in which 
chromodynamic--electroweak unification may be manifest is thus seen to remain an inadequacy of the SM landscape and therefore 
not to signal what could otherwise have been new development of it in respect of the relationship between a quark and a lepton. 
At LHC this conclusion has already been pushed fruitlessly to about three times as high a mass using time-like production and 
decay. The discovery at HERA was therefore that there was nothing new, which is as important as is any null result in science, the 
previous, most recent, example in particle physics being at LEP!

And finally, last but certainly not least HERA can close a new loop on the ever extending horizon of Dirac's original theoretical 
discovery concerning spin-$\frac{1}{2}$ fermions. Precision comparison of inelastic (posi)electron--proton scattering at the highest 
possible  with the expectations of point-like lepton plus point-like quark scattering, all now embedded in all that has been 
discovered concerning HERA's wealth of new understanding of hadronic structure and the validity of space-like GSW unification, 
leads to the firm statement that quarks are Dirac fermions at the distance scale of an attometer. It seems that nearly a hundred 
years on from the discovery of his equation, after it formed a template against which to establish the theoretical paradigm of field 
theory which underpins our SM, that at a distance scale of $\frac{1}{1000}$th the size of the atomic nucleus, $\frac{1}{100,000,000}$th 
of the size of an atom, we do understand what we are dealing with, namely still Dirac's fermions, even if we manifestly have still no 
understanding why they, with the GSW quantum fields which define their interactions, are there at all.

\subsection{Onwards?}

The significance and importance of the "HERA era" of physics can be seen as pivotal. For more than 24 years (1983--2007) the 
construction and operation of the first ever, and until now only, collider of different particle species probed all aspects of the 
Standard Model (SM). The impact of its scientific output has continued so far for a further 7 years. The combination of an energy 
reach beyond the Fermi scale, with the precision which the comprehensive experiments H1, ZEUS, HERMES and HERA-B brought 
to bear on the panoply of  physics, have provided a cornerstone in the development of the SM. Because only of the unforeseeable 
stubbornness of Nature, HERA has thus been able to secure with precision the consistency of the SM in the space-like domain, but 
without unfortunately opening up any understanding of the underlying reasons for the existence of the Dirac fermions, the leptons 
and the quarks, on which the SM is based.

This is nevertheless by any criterion a noble pedigree with an immensely important legacy. HERA's place is that of the contemporary 
manifestation by which has been discovered the nature and origin of matter since the discovery of the atomic nucleus. When taken 
alongside the pivotal experimental discovery by Thomson of his "corpuscle" in 1897 in Cambridge and the theoretical foundation of 
Dirac's original theory of the spin-$\frac{1}{2}$ electron, it can therefore be seen as that which has cemented in place the basis of the 
triumph of the latter half of 20th century physics, the Standard Model.

HERA's success has thus been pivotal. Further, and based on the success which is the SM, as we scramble to justify the future of 
particle physics in terms of any major new investment which costs factors more than an international initiative such as the ITER 
fusion facility, we must the humility not to forget the unchallengeable fact that nothing has ever been discovered beyond a Dirac 
fermion, and if it were to be discovered it would be truly revolutionary. Optimism is indeed essential, but with innovative pragmatism 
too! The way that HERA happened, the way it was funded with the HERA model by our predecessors, exemplified this innovative 
pragmatism. The way that HERA was designed, built and operated exemplified it. The excellence and importance of the science 
at HERA, which is so characteristic of the excellence of all who have had the privilege to work at DESY Hamburg, exemplifies it. 
In turn, we must exemplify it by continuing to recognise that none of it could have happened if the German taxpayers, along with 
their peers in all countries which have contributed to science at DESY, had not been willing to support our science following a 
reasonable request for financial investment. What we ask for next must exemplify it too. 

I thank everyone with whom I have had the privilege for decades to interact and work at DESY Hamburg, including my late wife 
Josephine whose contributions to DESY science I know from so many will also never be forgotten. My time at DESY will always 
be unforgettable doing great science. I also am sure that all who worked on HERA are indebted to the taxpayers of Germany, 
Europe and beyond who funded the HERA adventure. We hope like us that they all will agree that their investment in science at 
HERA at DESY has been priceless, and therefore certainly worth their investment!

%% file: acknowledgements.tex
\section*{Acknowledgements}

S. Pl\"{a}tzer acknowledges support through a Marie Curie Intra-European Fellowship under contract number PIEF-GA-2013-628739.  V. Myronenko, O. Turkot, K. Wichmann and M. Wing acknowledge the support 
of the Alexander von Humboldt Stiftung.
The work of F. Hautmann is supported in part by the DFG SFB 676 programme ``Particles, String and the Early Universe'' at the University of Hamburg and DESY.  
The work of E.R.~Nocera is supported by a STFC Rutherford Grant ST/M003787/1.  This activity was partially supported by the Israel Science Foundation.